\documentclass[aps,prd,twocolumn,preprintnumbers,superscriptaddress,tightenlines,
amsmath,amssymb,showpacs,nofootinbib]{revtex4-2}
\pdfoutput=1
\usepackage{textcomp}
\usepackage[english]{babel}
\usepackage{amsmath,amssymb,amsbsy,booktabs}
\usepackage{bm}
\usepackage{xcolor}
\usepackage{array}
\usepackage{amstext}
\usepackage{graphicx}
\usepackage{amsfonts}
\usepackage{bm}
\usepackage{dcolumn}
\usepackage{rotating}
\usepackage{epstopdf}
\usepackage{esint}
\usepackage{url}
\usepackage{pifont}
\usepackage{colortbl} 
\newcommand{\cmark}{\ding{52}}%
\newcommand{\xmark}{\ding{56}}%

\usepackage[colorlinks=true,
            linkcolor=blue,
            filecolor=blue,
            anchorcolor=blue,
            urlcolor=blue,
            citecolor=blue
            ]{hyperref}

\usepackage[utf8]{inputenc}

\usepackage[T1]{fontenc} 

\usepackage{multirow}

\usepackage{array}
\usepackage{booktabs}
\usepackage{times}
\allowdisplaybreaks[0]

\begin{document}

\title {\bf Scotogenic Dirac neutrino mass models embedded with leptoquarks:\\[0.15cm]
one pathway to address the flavor anomalies and the neutrino masses together}

\author{Shao-Ping Li}
\email{ShowpingLee@mails.ccnu.edu.cn}
\affiliation{Institute of Particle Physics and Key Laboratory of Quark and Lepton Physics~(MOE),\\
	Central China Normal University, Wuhan, Hubei 430079, China}

\author{Xin-Qiang Li}
\email{xqli@mail.ccnu.edu.cn}
\affiliation{Institute of Particle Physics and Key Laboratory of Quark and Lepton Physics~(MOE),\\
	Central China Normal University, Wuhan, Hubei 430079, China}

\author{Xin-Shuai Yan}
\email{xinshuai@mail.ccnu.edu.cn}
\affiliation{Institute of Particle Physics and Key Laboratory of Quark and Lepton Physics~(MOE),\\
	Central China Normal University, Wuhan, Hubei 430079, China}

\author{Ya-Dong Yang}
\email{yangyd@mail.ccnu.edu.cn}
\affiliation{Institute of Particle Physics and Key Laboratory of Quark and Lepton Physics~(MOE),\\
	Central China Normal University, Wuhan, Hubei 430079, China}
\affiliation{Institute of Particle and Nuclear Physics, Henan Normal University, Xinxiang, Henan 453007, China}

\begin{abstract}
If the leptoquarks proposed to account for the intriguing anomalies observed in the semileptonic $B$-meson decays, $R_{D^{(\ast)}}$ and $R_{K^{(\ast)}}$, as well as in the anomalous magnetic moment of the muon, $(g-2)_\mu$, can be embedded into the scotogenic Dirac neutrino mass models, all these flavor anomalies, together with the origin of neutrino masses and the nature of dark matter, could be potentially addressed in a unified picture. Among the minimal seesaw, one-loop, and two-loop realizations of the dimension-4 effective operator $\mathcal{L}_{4}$ for the Dirac neutrino masses, we show that plenty of diagrams associated with the two-loop realizations of $\mathcal{L}_{4}$ can support the coexistence of leptoquarks and dark matter candidates. After a simple match of these leptoquarks with the ones introduced to accommodate all the flavor anomalies, we establish the scotogenic Dirac neutrino mass models embedded with leptoquarks, which could address all the problems mentioned above.

\end{abstract}


\maketitle

\section{Introduction} 

Despite its great success in elementary particle physics, the Standard Model (SM) fails to explain the origin of tiny neutrino masses and the nature of dark matter (DM). In general, these two problems are considered as two separate topics, and could be solved via completely different mechanisms with unrelated particles and/or interactions. However, addressing them in a unified picture, in which the neutrino mass scale as well as the DM property and its abundance can be quantitatively connected with each other, would be more intriguing. It is known that the scotogenic models can fulfill the task, because the essence of these models, as demonstrated in the original works~\cite{Ma:2006km,Kubo:2006yx}, is that the neutrino mass generation involves at least one DM propagator. Interestingly enough, the interactions responsible for the neutrino mass generation, though being not necessarily, can also successfully account for the DM relic abundance~\cite{Farzan:2012sa}. Thus, in a sense, the scotogenic models could really kill two birds with one stone.

Depending on the nature of neutrinos, being of the Majorana or of the Dirac type, the scotogenic models are generally classified into two categories. Historically, neutrinos of the Majorana type are relatively more motivated and various mechanisms for their mass generation, based either on the seesaw~\cite{Minkowski:1977sc,Yanagida:1979as,Gell-Mann:1979vob,Mohapatra:1979ia,Schechter:1980gr,Schechter:1981cv,Foot:1988aq} or the loop effects~\cite{Zee:1980ai,Zee:1985id,Babu:1988ki,Ma:2006km}, have been proposed. However, since no indisputable evidence has been reported for the neutrinoless double beta decay so far~\cite{EXO-200:2014ofj,KamLAND-Zen:2016pfg,NEMO-3:2016qxo,Agostini:2017iyd,CUORE:2017tlq}, the nature of neutrinos is still unsettled, and the possibility of Dirac neutrinos should not be discounted. In fact, during the past few years there has been a renewed interest in building the mass generation models for the Dirac neutrinos, in which the tiny masses can be generated either through the seesaw mechanisms~\cite{Roncadelli:1983ty,Ma:2014qra,Ma:2015mjd,Ma:2015raa,Valle:2016kyz,CentellesChulia:2016rms,CentellesChulia:2016fxr,Reig:2016ewy,CentellesChulia:2017koy,CentellesChulia:2017sgj,Borah:2017leo,Bonilla:2017ekt,Borah:2017dmk,Borah:2018nvu,Ma:2018bow,Borah:2019bdi,Gu:2006dc,Gu:2016hxh} or the loop effects~\cite{Farzan:2012sa,Okada:2014vla,Bonilla:2016diq,Wang:2016lve,Ma:2017kgb,Wang:2017mcy,Helo:2018bgb,Reig:2018mdk,Han:2018zcn,Kang:2018lyy,Bonilla:2018ynb,Calle:2018ovc,Carvajal:2018ohk,Ma:2019yfo,Bolton:2019bou,Saad:2019bqf,Bonilla:2019hfb,Dasgupta:2019rmf,Jana:2019mez,Enomoto:2019mzl,Ma:2019byo,Restrepo:2019soi}. Among these proposals, if the DM exchanges are also involved, they can be identified as the scotogenic Dirac neutrino mass models (SD$\nu$M)~\cite{Farzan:2012sa}, which constitutes one of the key ``stones'' of this work. Besides in the mass generation mechanisms, growing interests have also been revived in other aspects of the Dirac neutrinos, such as their intimate connections to the baryon asymmetry of the Universe~\cite{Li:2020ner,Li:2021tlv}, which is another conundrum the SM of particle physics is now facing.

Besides the problems associated with neutrinos and DM, several intriguing anomalies in the semileptonic $B$-meson decays, particularly in the ratios $R_{D^{(\ast)}}$~\cite{BaBar:2012obs,BaBar:2013mob,Belle:2015qfa,LHCb:2015gmp,Belle:2016dyj,Belle:2017ilt,LHCb:2017smo,LHCb:2017rln,Belle:2019rba} and $R_{K^{(\ast)}}$~\cite{LHCb:2014vgu,LHCb:2017avl,Belle:2019oag,BELLE:2019xld,LHCb:2021trn,LHCb:2021lvy}, as well as in the muon $g-2$ measurements~\cite{Muong-2:2006rrc,Muong-2:2021ojo} have been reported over the last few years. Interestingly, it has been demonstrated that models with only a single leptoquark (LQ)~\cite{Du:2021zkq,Ban:2021tos,Cheung:2022zsb} or a few of them~\cite{Chen:2017hir,Bigaran:2019bqv,Crivellin:2019dwb,Saad:2020ihm,Babu:2020hun,Lee:2021jdr,Julio:2022bue,DAlise:2022ypp} are capable of addressing all these flavor anomalies simultaneously. Inspired by the spirit of SD$\nu$M, we will explore in this work if these ``stones'', once combined properly, could accommodate all the flavor anomalies, together with the origin of neutrino masses and the nature of dark matter in a unified picture. Arguably, the simplest approach is to check if whatever proposed to account for the $B$-meson and $(g-2)_{\mu}$ anomalies simultaneously also contributes to the Dirac neutrino mass generation. If these viable LQs could be embedded into the SD$\nu$M, not only can all the aforementioned problems be solved in a unified picture, but also their rich phenomenology correlated with the neutrino mass and mixing will make the constructed models more predictable and testable.    

Our starting point will be the classification of minimal seesaw (tree-level), one-loop, and two-loop realizations of the Dirac neutrino mass operators $\mathcal{L}_{4}$ at dimension four~\cite{Ma:2016mwh,CentellesChulia:2019xky}.\footnote{Note that the classification of seesaw and loop realizations of the Dirac neutrino mass operators at dimension five~\cite{Yao:2018ekp,CentellesChulia:2018gwr,Jana:2019mgj} and dimension six~\cite{Yao:2017vtm,CentellesChulia:2018bkz}
also exists; our choice of $\mathcal{L}_{4}$ here is solely for simplicity} Guided exclusively by the SM gauge symmetry, we will work out a selection scheme for the topologies, and then examine systematically possible diagrams and ultraviolet (UV) completions associated with each topology. In contrast to the previous studies~\cite{Ma:2016mwh,CentellesChulia:2019xky}, the topologies we are seeking must involve the exchanges of at least one LQ and one DM candidate, supporting therefore our scenario of SD$\nu$M embedded with LQs---only after the dominance of their contributions to the Dirac neutrino masses is established, will the topology-based UV completions be dubbed LQ-SD$\nu$M. If the LQ(s) embedded can simultaneously account for the $B$-meson and $(g-2)_{\mu}$ anomalies as well, the LQ-SD$\nu$M will be considered as the mighty ``stones''. As will be demonstrated in this work, there exist plenty of such kinds of mighty models. Intriguingly, in some of these models, a close-knit connection can be established between the flavor anomalies and the neutrino mass.
  
The paper is organized as follows. We begin Sec.~\ref{sec:classfication} by establishing a topology-selection scheme, under which we examine the topologies and their associated diagrams of the two-loop realizations of $\mathcal{L}_{4}$, and identify the ones that can support the coexistence of at least one LQ and one DM candidate. Based on the existing studies aimed at addressing all the flavor anomalies with LQs exclusively, we establish the possible mighty ``stones'' in Sec.~\ref{sec:mighty_stones}. Our conclusions are finally made in Sec.~\ref{sec:con}. For convenience, useful supplementary materials are provided in the appendices. 
  
\section{\boldmath SD$\nu$M embedded with LQs}
\label{sec:classfication}

In order to have massive Dirac neutrinos, one needs firstly extend the SM particle contents by adding the right-handed neutrinos $\nu_R$. Arguably, the simplest way goes with the following Yukawa interaction, 
\begin{align}
\mathcal{L}_{4}=-y\bar{L}_L\widetilde{H}\nu_R+\text{H.c.},\label{eq:SM_yukawa}
\end{align}
where $L_L=(\nu_L,e_L)^T$ is the left-handed lepton doublet, and $\widetilde{H}=i\sigma_2H^*$ with $\sigma_2$ the second Pauli matrix and $H=(\phi^+,\phi^0)^T$ the SM Higgs doublet. However, to account for the sub-eV neutrino masses, the Yukawa coupling $y$ must be tuned to a very small value, $y\sim \mathcal{O}(10^{-12})$, being therefore often criticized as \textit{unnatural}. To circumvent this problem, more attractive mechanisms, such as the Dirac seesaw~\cite{Roncadelli:1983ty,Gu:2006dc,Ma:2014qra,Ma:2015mjd,Ma:2015raa,Valle:2016kyz,CentellesChulia:2016rms,CentellesChulia:2016fxr,Reig:2016ewy,Gu:2016hxh,CentellesChulia:2017koy,CentellesChulia:2017sgj,Borah:2017leo,Bonilla:2017ekt,Borah:2017dmk,Borah:2018nvu,Ma:2018bow,Borah:2019bdi} and the radiative mass generation~\cite{Farzan:2012sa,Okada:2014vla,Bonilla:2016diq,Wang:2016lve,Ma:2017kgb,Wang:2017mcy,Helo:2018bgb,Reig:2018mdk,Han:2018zcn,Kang:2018lyy,Bonilla:2018ynb,Calle:2018ovc,Carvajal:2018ohk,Ma:2019yfo,Bolton:2019bou,Saad:2019bqf,Bonilla:2019hfb,Dasgupta:2019rmf,Jana:2019mez,Enomoto:2019mzl,Ma:2019byo,Restrepo:2019soi}, have been proposed, and they are not necessarily confined to the Yukawa structure specified by Eq.~\eqref{eq:SM_yukawa}. 

In this paper, we will focus on the case in which the Dirac neutrino masses are generated by these attractive mechanisms. In addition, a global lepton-number symmetry $\text{U}(1)_L$ will be introduced to ensure the absence of Majorana mass terms for $\nu_R$ at all orders, so that the Dirac nature of the neutrinos is protected. In general, there might be an infinite number of UV completions of $\mathcal{L}_{4}$ that respect the SM and lepton-number symmetries. For simplicity, we will only focus on the fields transforming as singlet, doublet, or triplet under the SM $\text{SU}(2)_{\text{L}}$ gauge symmetry. Furthermore, for the sake of minimality, we will introduce only limited number of new degrees of freedom carrying colors.
Before diving into the classification of minimal seesaw (tree-level), one-loop, and two-loop realizations of $\mathcal{L}_{4}$, and identifying the topology and its associated diagrams that can support the coexistence of at least one LQ and one DM candidate, we will firstly give a brief discussion about the LQs and DM candidates, and then establish a topology-selection scheme. 

\begin{table}[t]
	\begin{center}	
		\renewcommand*{\arraystretch}{1.5}
		\tabcolsep=0.29cm
		\begin{tabular}{clcl}
			\hline \hline
			Scalar LQ & SM Rep. & Vector LQ & SM Rep.
			\\ \hline \rowcolor{lightgray}
			\begin{tabular}{@{}c@{}}
				$S_1 \bar{Q}^C_L L_L$ \\ $S_1\bar{u}^C_Re_R$ \\  $S_1\bar{d}^C_R\nu_R$
			\end{tabular}
			& $(\bar{3}, 1, 1/3)$ & 
			\begin{tabular}{@{}c@{}}
				$U_{1\mu}\bar{Q}_L\gamma^{\mu}L_L$ \\ $U_{1\mu}\bar{d}_R\gamma^{\mu}e_R$ \\ $U_{1\mu}\bar{u}_R\gamma^{\mu}\nu_R$
			\end{tabular}
		    & $(3, 1, 2/3)$\\ 
		    $S_3 \bar{Q}^C_L L_L$ & $(\bar{3}, 3, 1/3)$ & $U_{3\mu} \bar{Q}_L\gamma^{\mu}L_L$ & $(3, 3, 2/3)$ \\  \rowcolor{lightgray}
		    \begin{tabular}{@{}c@{}}
		    	$R_2\bar{u}_R L_L$ \\ $R_2\bar{Q}_L e_R $
		    \end{tabular}
			 & $(3, 2, 7/6)$ & 
			 \begin{tabular}{@{}c@{}}
			 	$V_{2\mu}\bar{d}^C_R\gamma^{\mu} L_L$ \\   $V_{2\mu}\bar{Q}^C_L\gamma^{\mu}e_R$
			 \end{tabular}
		  & $(\bar{3}, 2, 5/6)$\\ 
			$\widetilde{S}_1 \bar{d}^C_R e_R$ & $(\bar{3}, 1, 4/3)$ & $\widetilde{U}_{1\mu} \bar{u}_R\gamma^{\mu}e_R$ & $(3, 1, 5/3)$ \\ \rowcolor{lightgray}
			\begin{tabular}{@{}c@{}}
				$\widetilde{R}_2 \bar{d}_R L_L$ \\$\widetilde{R}_2 \bar{Q}_L \nu_R$
			\end{tabular}
			 & $(3, 2, 1/6)$ & 
			 \begin{tabular}{@{}c@{}}
			 	$\widetilde{V}_{2\mu}\bar{u}^C_R\gamma^{\mu}L_L$  \\ $\widetilde{V}_{2\mu}\bar{Q}^C_L\gamma^{\mu}\nu_R$ 
			 \end{tabular} & $(\bar{3}, 2, -1/6)$	\\ 
			$\bar{S}_1 \bar{u}^C_R\nu_R $ & $(\bar{3}, 1, -2/3)$ & $\bar{U}_{1\mu}\bar{d}\gamma^{\mu}\nu_R$ & $(3, 1, -1/3)$\\
			\hline\hline
		\end{tabular}
    	\caption{Possible couplings of the scalar and vector LQs to the SM fermions and the right-handed neutrinos $\nu_R$, as well as their representations under the SM gauge group $\text{SU(3)}_{c}\otimes \text{SU}(2)_{\text{L}}\otimes \text{U}(1)_{\alpha}$. The interactions between LQs and fermions are indicated only schematically and, for simplicity, their coupling constants and the Hermitian conjugated terms are not shown explicitly---they could be found, e.g., in Ref.~\cite{Dorsner:2016wpm}. Our convention for the hypercharge $\alpha$ is specified by $Q_{\text{em}}=T_3+\alpha$, and the charge conjugate of a fermion field $\psi$ is defined as $\psi^C=C\bar{\psi}^T$, with $C$ the charge conjugate operator.} 
    	\label{tab:LQ}
	\end{center} 
    \vspace*{-0.45cm}
\end{table}

The LQs, due to their ability of turning quarks into leptons and vice versa, have very rich phenomenology in, e.g., the anomalous magnetic moment of the charged leptons, the weak decays of various hadrons, the neutral-meson mixings, etc.; see Ref.~\cite{Dorsner:2016wpm} for a recent review. If their interactions with the right-handed Dirac neutrinos $\nu_R$ are taken into account, there are totally twelve LQs, namely six scalars and six vectors. Their possible couplings to the SM fermions and $\nu_R$, as well as their representations under the SM gauge symmetry $\text{SU(3)}_{c}\otimes \text{SU}(2)_{\text{L}}\otimes \text{U}(1)_{\alpha}$ are summarized in Table~\ref{tab:LQ}. We follow here the same conventions as used in Ref.~\cite{Dorsner:2016wpm}.

Mounting evidences have indicated that the DM, if being an elementary particle, must be stable, colorless, and electrically neutral. To avoid the constraints from the DM direct detection experiments, such as PandaX-II~\cite{PandaX-II:2016wea} and XENON1T~\cite{XENON:2017vdw}, we will, as discussed in Refs.~\cite{Yao:2018ekp,Yao:2017vtm}, require the DM candidate to satisfy the condition $\alpha=0$, with $\alpha$ the hypercharge of the DM field, since otherwise the direct detection cross section via nucleon recoil, being proportional to $\alpha^2$, would be generally quite large. Consequently, the convention $Q_{\text{em}}=T_3+\alpha$, together with the requirement $Q_{\text{em}}=0$, leads further to the condition $T_3=0$ for the DM candidate. This implies that the $\text{SU}(2)_{\text{L}}$ multiplets with even number of components (i.e., doublet, quartet, etc.) are already eliminated. The scalar doublet with $\alpha=\pm 1/2$ is, however, an exception, because a mass splitting enforced between the scalar and pseudo-scalar components can eliminate the coupling of the DM candidate to the $Z$ boson at tree level~\cite{LopezHonorez:2006gr,Farzan:2012ev}, and thus help the model with such a scalar doublet evade the constraints from the direct detection experiments. In general, there can be more than one DM candidate participating in the neutrino mass generation. If so, the lightest one will be deemed the DM.  

To prevent the DM from decaying exclusively to the SM particles, a necessary auxiliary symmetry is usually introduced. The most popular choice is the $Z_2$ symmetry, which is commonly used in radiative neutrino mass models to help stabilize the DM candidates. Assuming the symmetry to be exact, we will assign an even ($+$) $Z_2$ parity to all the SM particles, as well as the right-handed neutrinos $\nu_R$, whereas an odd ($-$) $Z_2$ parity to the DM candidates. 

Summarizing all the bits and pieces of our discussions made above, we can establish the following set of criteria for our later topology (diagram) selection scheme:
\begin{itemize}
	\item Only the fields transforming as singlet, doublet, or triplet under the $\text{SU}(2)_{\text{L}}$ gauge symmetry will be considered.
	
	\item The DM candidates, expect for being the neutral components of a scalar doublet with  $\alpha=\pm 1/2$, must satisfy the condition $\alpha=T_3=0$, to avoid the constraints from DM direct detection experiments.
	
	\item A minimum of new degrees of freedom carrying colors will be introduced.
		
	\item The dark $Z_2$ symmetry introduced to prevent the DM from decaying exclusively to the SM particles must be exact.
\end{itemize} 

\begin{figure}[t]
	\centering
	\includegraphics[width=0.48\textwidth]{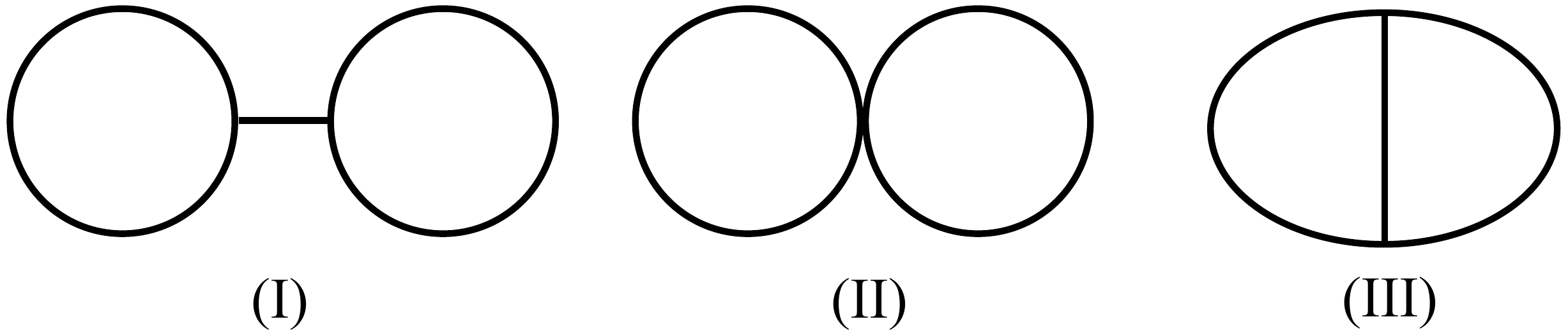}
	\caption{Three possible structures for a topology (diagram) containing at least one LQ and one DM at the two-loop level. Note that the particle connecting the two loops in structure (I) must be colorless.} 
	\label{fig:loop} 
\end{figure}

Given that the external field contents $L_L$, $\nu_R$, and $H$ in $\mathcal{L}_{4}$ are all colorless, the requirement of $\text{SU(3)}_{c}$ symmetry indicates that the colored internal fields must be wrapped into loops. If they all arise at one loop, it is obvious that the DM candidate cannot appear in the very same loop, because it must be colorless. Then, in order to have a diagram containing at least one LQ and one DM, the DM must be linked to the colored loop (i.e., the one containing the LQ) either directly or through a colorless portal. In either case, since the DM possesses a $Z_2$-odd parity, whereas all the external field contents are even under the $Z_2$ symmetry, the DM must also be wrapped into loops to preserve the $Z_2$ symmetry. Furthermore, if the DM forms a separate loop with other colorless, $Z_2$-odd particles, our desired diagrams must have either the structure (I) or the structure (II) depicted in Fig.~\ref{fig:loop}, because the maximum loop number concerned here is restricted to two; if the DM appears in a loop that shares a common piece with the colored loop, the desired diagrams would be of the structure (III) shown in Fig.~\ref{fig:loop}. On the other hand, if the DM and the colored fields share a loop, the diverted colors from this loop have to be shunted into another loop. Once again, the diagrams in this case would possess the structure (III). Finally, the topology (diagram) of the structure (I) is one-particle reducible, and thus should be discarded when discussing the loop realizations of an effective neutrino mass operator, because the line which would disconnect the diagram must have the quantum number to mediate the seesaw realizations of the same mass operator~\cite{Cepedello:2018rfh}. As a consequence, our analyses will focus only on the two-loop realizations of the operator $\mathcal{L}_{4}$ with the last two structures depicted in Fig.~\ref{fig:loop}.

\subsection{\boldmath Two-loop realizations of the operator $\mathcal{L}_{4}$}

\begin{figure}[t]
	\centering
	\includegraphics[width=0.48\textwidth]{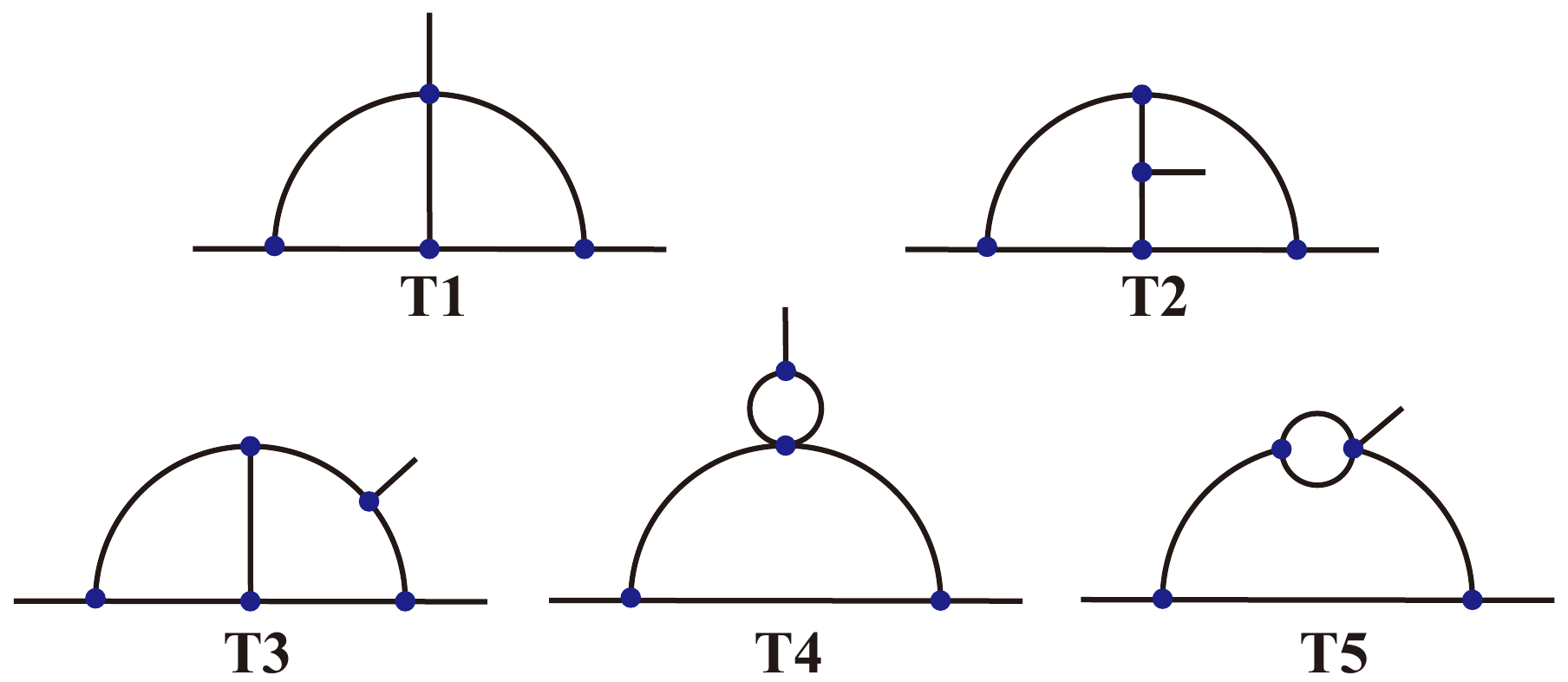}
	\caption{Finite two-loop 1PI topologies with three- and four-point vertices and three external legs, which are all associated with the two-loop realizations of the operator $\mathcal{L}_{4}$~\cite{CentellesChulia:2019xky}.} 
	\label{fig:4d2looptop} 
\end{figure} 

Two-loop realizations of the effective operator $\mathcal{L}_{4}$ have been studied in Ref.~\cite{CentellesChulia:2019xky}. After removing all the topologies corresponding to the tadpoles, the self-energy diagrams, and the non-renormalizable diagrams involving, e.g., the three-point vertices with only fermions or the four-point vertices with a fermion insertion, we are left with five one-particle-irreducible (1PI) topologies, which are shown in Fig.~\ref{fig:4d2looptop}. It can be seen that the topologies T1, T2, T3, and T5 all share the structure (III), while T4 has the structure (II). 

\begin{figure}[t]
	\centering
	\includegraphics[width=0.48\textwidth]{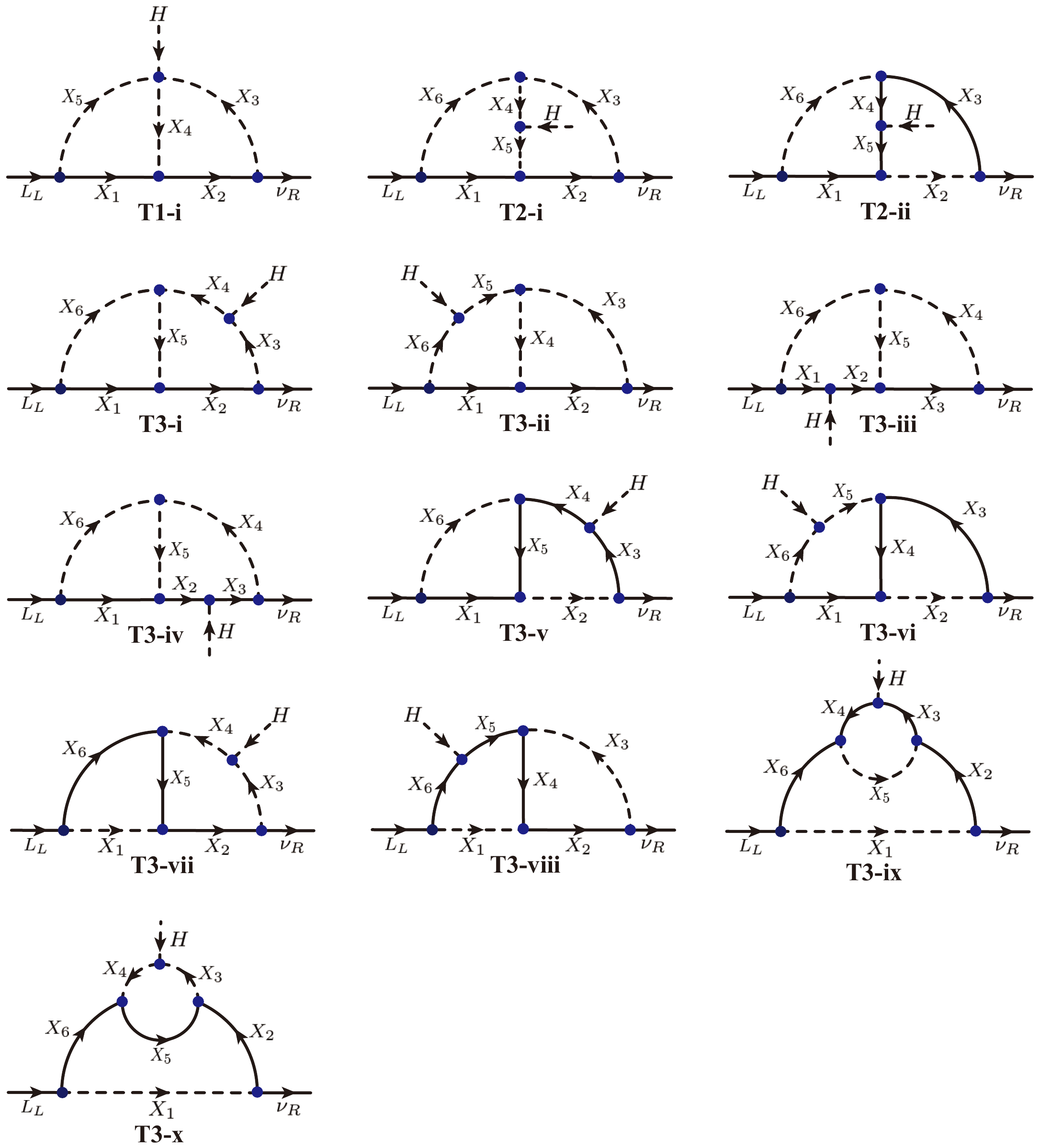}
	\caption{Genuine two-loop diagrams (the top three) and the ones containing a compressible fermion-fermion-scalar vertex~\cite{CentellesChulia:2019xky}. Here the dashed line represents either a scalar or a vector field, while the solid line always denotes a fermion field.}
	\label{fig:4d2loop} 
\end{figure} 

\begin{figure}[t]
	\centering
	\includegraphics[width=0.48\textwidth]{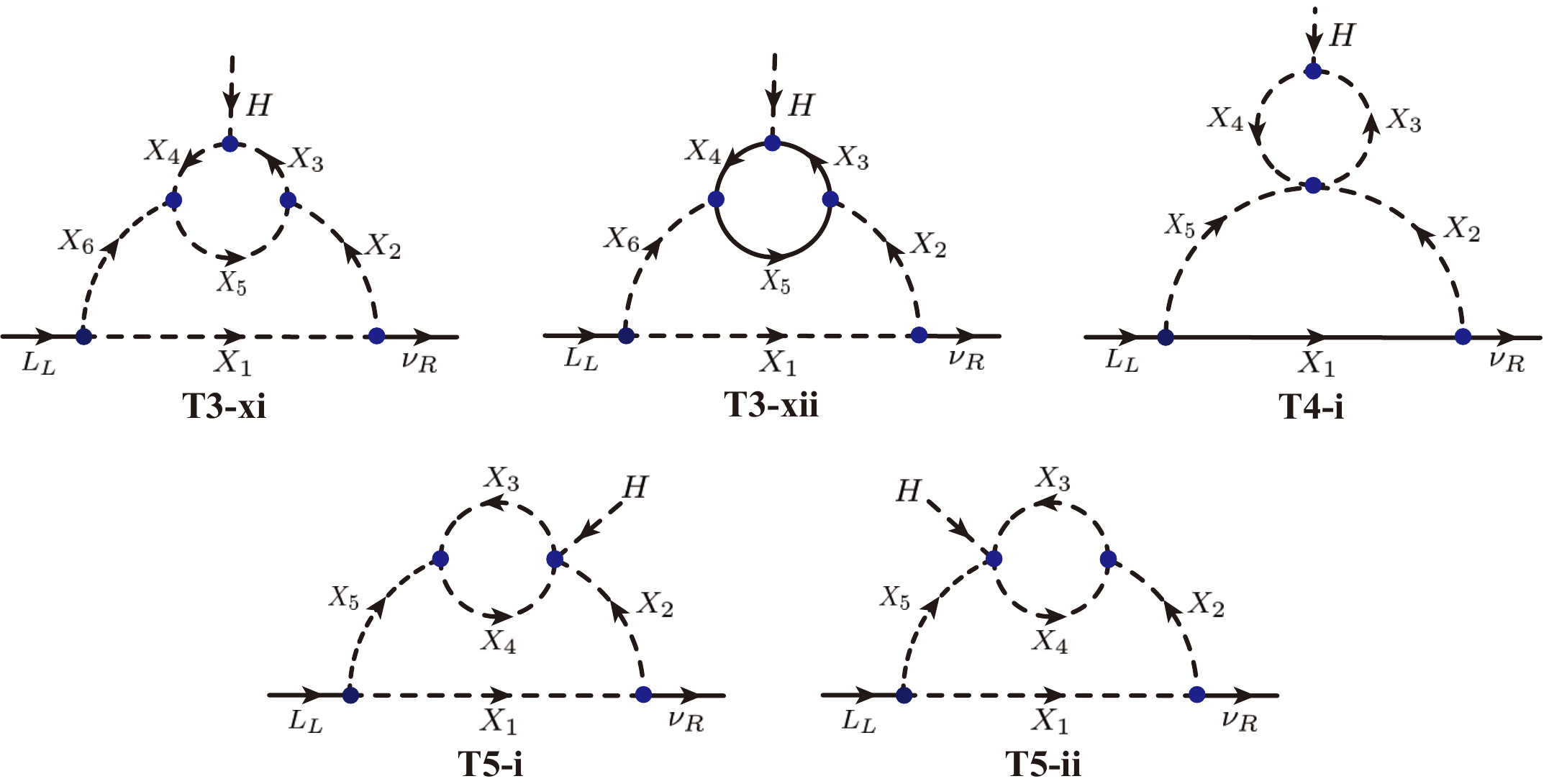}
	\caption{Two-loop diagrams that contain a compressible scalar-scalar-scalar or a compressible scalar-vector-vector vertex. The other captions are the same as in Fig.~\ref{fig:4d2loop}.}
	\label{fig:4d2delet} 
\end{figure} 

Based on these five topologies, we can build 18 diagrams after specifying the external fermion and scalar fields, which are depicted in Figs.~\ref{fig:4d2loop} and \ref{fig:4d2delet}. It can be seen that all the diagrams in Fig.~\ref{fig:4d2loop}, except for the top three, contain a compressible fermion-fermion-scalar vertex, whereas the diagrams in Fig.~\ref{fig:4d2delet} have a compressible scalar-scalar-scalar or a compressible scalar-vector-vector vertex. As the top three diagrams in Fig.~\ref{fig:4d2loop} contain no compressible three- or four-point renormalizable vertex, they will be called the genuine two-loop diagrams~\cite{CentellesChulia:2019xky}. Here the compressibility means that an internal sub-loop can be compressed to a renormalizable vertex, while the genuinity of a diagram indicates that its contribution to the neutrino masses arises firstly at the two-loop level~\cite{CentellesChulia:2019xky}. 

\begin{figure}[t]
	\centering
	\includegraphics[width=0.48\textwidth]{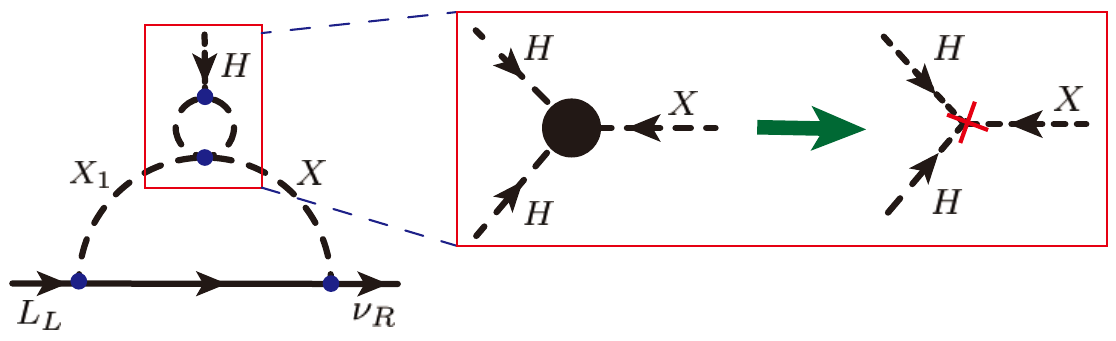}
	\caption{The sub-diagram surrounded by the small red box on the left is a compressible scalar-scalar-scalar (scalar-vector-vector) loop, and always generates a tree-level three-point vertex $HXX_1$, unless $X_1=H$ and $X$ transforms as $(1,1,-1)$ under the SM gauge group, because the local operator $H(x)H(x)X(x)$ (on the right inside the big red box) vanishes automatically due to the anti-symmetric contraction of the two $\text{SU}(2)_{\text{L}}$ doublets $H$ to the singlet $X$, whereas the non-local operator $H(x)H(y)X(z)$ realized at one loop (on the left inside the big red box) does not vanish in general~\cite{Cepedello:2018rfh,CentellesChulia:2019xky}.}
	\label{fig:F4} 
\end{figure}

Let us firstly consider the diagrams in Fig.~\ref{fig:4d2delet}. For a simple demonstration, here we will focus on the diagram T4-i that has the structure (II). If the DM candidates occupy the upper, small loop, then, to make this diagram our desired one, at least one LQ must propagate in the lower, big loop, yielding therefore a compressible three-point vertex $X_2X_5H$ with both $X_{2}$ and $X_{5}$ carrying colors. On the other hand, If the LQ arises in the upper, small loop, then the DM candidates must propagate in the lower, big loop, resulting in another compressible three-point vertex $X_2X_5H$ with both $X_{2}$ and $X_{5}$ being colorless and carrying a $Z_2$-odd parity. Nevertheless, the resulting non-vanishing three-point vertices in both cases, according to Refs.~\cite{Cepedello:2018rfh,CentellesChulia:2019xky} and also shown in Fig.~\ref{fig:F4}, cannot prevent the presence of tree-level vertices $X_2X_5H$, or do not satisfy our requirement on the fields $X_{2}$ and $X_{5}$ for the non-local operator $H(x)H(y)X(z)$ realized in Fig.~\ref{fig:F4}, rendering therefore the two-loop diagram non-genuine~\cite{CentellesChulia:2019xky}. Consequently, we will consider neither the diagram T4-i nor the rest ones in Fig.~\ref{fig:4d2delet} (due to the same reason), and thus the structure (II) in Fig.~\ref{fig:loop} is completely eliminated. 

\begin{figure}[t]
	\centering
	\includegraphics[width=0.48\textwidth]{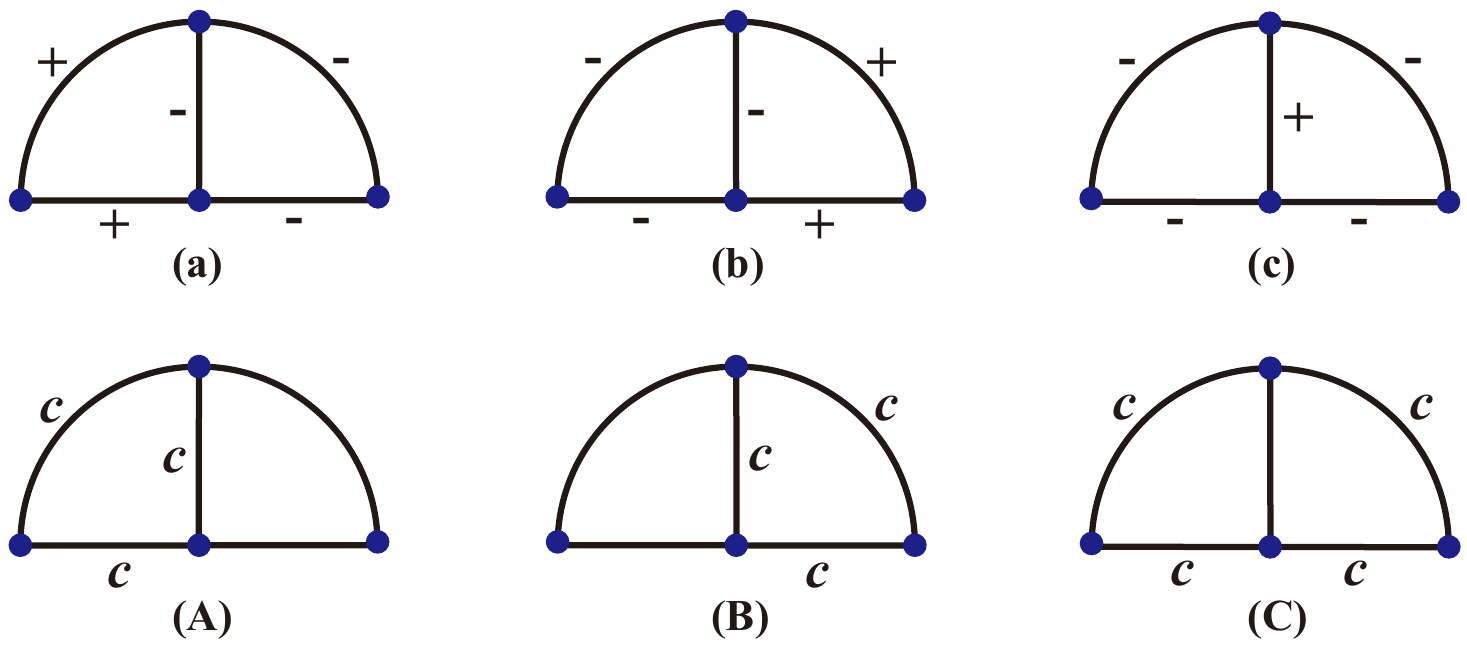}
	\caption{Possible assignments of the $Z_2$ parity (the upper three diagrams) and the color $c$ (the lower three diagrams) to the internal fields of the structure diagram. Note that the diagrams with all their internal fields labeled with $c$ or $+$ have been eliminated, because no place is left for the DM candidates in this case.}
	\label{fig:Z2} 
\end{figure} 

To build the possible UV models associated with the diagrams shown in Fig.~\ref{fig:4d2loop}, we have to pinpoint the proper quantum numbers of all the internal fields in these diagrams. Let us firstly assign them possible $Z_2$ parity and color. Note that assigning concrete $\text{SU}(3)_c$ representations to the internal fields is not necessary at this point; we will thus use a simple notation ``$c$'' to indicate that the corresponding fields are colored. In addition, since these diagrams share the same structure, we can strip away the external fields to make our following procedure as general as possible. Then, it can be seen from Fig.~\ref{fig:Z2} that there are three ways to assign the $Z_2$ parity (the upper three diagrams in Fig.~\ref{fig:Z2}) and the color $c$ (the lower three diagrams in Fig.~\ref{fig:Z2}) to the internal fields separately, resulting in nine combinations in total. Among them, the combinations (a)-(B) and (b)-(A) leave no place for the DM candidates, and hence should be eliminated. In addition, the combinations (a)-(C), (b)-(C), (c)-(A), and (c)-(B) contain two colored, $Z_2$-odd particles, whereas (c)-(C) involves four colored, $Z_2$-odd particles. They have to be eliminated as well according to our selection criteria. Finally, the combinations (a)-(A) and (b)-(B) contain one colored, $Z_2$-odd particle, two colored, $Z_2$-even particles (one of them can be a LQ), and two colorless, $Z_2$-odd particles (one of them can be a DM). Therefore, these two combinations are what we will assign to the diagrams in Fig.~\ref{fig:4d2loop}. For the convenience of our later discussions, we will hereafter use ``a'' and ``b'' to denote the combinations (a)-(A) and (b)-(B), respectively. 

Assigning the LQs in Table~\ref{tab:LQ} to the internal field contents for both the combinations ``a'' and ``b'', we work out in Tables~\ref{tab:T1-i-a}--\ref{tab:T3-x-a} all the possible UV models corresponding to the diagrams shown in Fig.~\ref{fig:4d2loop}, except for T2-ii, T3-v, T3-vi, T3-vii, and T3-viii, since their UV models can be obtained by following the same procedure (note that they all involve at least one colored, $Z_2$-odd fermion). In addition, the possible DM candidates are already identified by using our prescriptions for the quantum numbers of them. As indicated by the quantum numbers of the field contents in Tables~\ref{tab:T3-ix-a} and \ref{tab:T3-x-a}, we have set the LQs and the DM candidates, which occupy the lower and the upper loop respectively, to be the combination ``a'' for the diagrams T3-ix and T3-x. The underlying reason for such an arrangement is to  
make the LQs' couplings to the left-handed lepton doublet $L_L$ contribute directly to the 
neutrino mass generation, which in turn helps establish the mighty models to be introduced later, because these interactions are essential to address the flavor anomalies, as will be discussed in Sec.~\ref{sec:mighty_stones}. This indicates that the combination ``b'' in these two cases, i.e., the LQs and the DM candidates occupying the two loops in a reversed order, must be eliminated, since otherwise no LQ couplings in Table~\ref{tab:LQ} will contribute to the neutrino mass generation. It is also interesting to note that only a handful of UV models can involve a pair of LQs listed in Table~\ref{tab:LQ}. They are ($\widetilde{R}_2$, $\bar{S}_1$) and ($\widetilde{V}_2$, $\bar{U}_1$) for the diagram T3-i with the combination ``b'', as well as ($\widetilde{R}_2$, $S_{1,3}$), ($R_2$, $\bar{S}_1$), ($\widetilde{V}_2$, $U_{1,3}$), and ($V_2$, $\bar{U}_1$) for the diagram T3-ii with ``a''.

These UV models for radiative Dirac neutrino masses support the coexistence of the LQs and the DM candidates, and can be easily read off from Tables~\ref{tab:T1-i-a}--\ref{tab:T3-x-a}. Let us take the model T1-i-a-A-1 as an illustration.\footnote{Here T1-i-a denotes the diagram T1-i depicted in Fig.~\ref{fig:4d2loop} and with the combination (a)-(A) shown in Fig.~\ref{fig:Z2}, and T1-i-a-A-1 the model listed in the first row of Table~\ref{tab:T1-i-a}. We refer the readers to Appendix~\ref{sec:appB} for our convention of the model labeling.} The new field contents of this model consist of two SM singlets $X^F_2$ and $X^S_3$, one colored, $\text{SU}(2)_{\text{L}}$ doublet $X^S_4$, and one LQ $S_1$. Note that, for simplicity, we have chosen $X_{3,4}$ to be the scalar fields---hence the superscript $S$ on $X_{3,4}$ here, though they can also be vector ones, and $X_{2,3,5}$ to be the $\text{SU}(2)_{\text{L}}$ singlets, even though triplets would work as well. After setting the hypercharge $\alpha=0$ for $X_1$, both $X_{2}$ and $X_{3}$, as shown in Table~\ref{tab:T1-i-a},  
satisfy our prescription for the DM candidates, i.e., carrying the hypercharge $\alpha=0$, and thus the lighter one shall be the DM. 

With our choice for the field contents, the relevant Lagrangian that generates the diagram T1-i-a is then given by 
\begin{align}\label{eq:sample}
	\mathcal{L}\supset& [\lambda_{1}\bar{Q}_L^{C}i\tau_2L_{L}S_1+\lambda_{2}\bar{Q}_LX_{2L}^CX_4
	+\lambda_{3}\bar{\nu}_RX_{2L}X^{\dagger}_3\nonumber \\[0.12cm]
	& +\lambda_{4}X^{\dagger}_4HS^{\dagger}_1X_3-M_{X_2}\bar{X}_{2L}X_{2R}+\text{H.c.}] \nonumber \\[0.12cm]
	&-M^2_{S_1}S_1^{\dagger}S_1-M^2_{X_3}X_3^{\dagger}X_3-M^2_{X_4}X_4^{\dagger}X_4,
\end{align}
where $X_{2}$ is a vector-like fermion, $\lambda_{1,2,3}$ describe the new Yukawa interactions, and $\lambda_{4}$ is the coupling constant among the Higgs $H$, the LQ $S_1$, and the two other new scalar fields $X_{3,4}$. After the electroweak symmetry breaking, one can obtain the effective neutrino mass by computing the two-loop integral of this diagram. For technical details and the final expression of the loop evaluation, we refer the readers to Refs.~\cite{vanderBij:1983bw,McDonald:2003zj,Angel:2013hla,AristizabalSierra:2014wal}. Following the same procedure, one can also write out the Lagrangians for all the other models and derive the corresponding neutrino masses.

\subsection{Securing the two-loop dominance}
\label{sec:LD}

To ensure the dominance of the two-loop contribution to the Dirac neutrino masses, all the lower-order corrections must be forbidden. For instance, besides the renormalizable Yukawa interaction $\bar{L}_L\widetilde{H}\nu_R$, all the tree-level (seesaw) and one-loop realizations of the effective operators $\mathcal{L}_{4}$ considered, e.g., in Ref.~\cite{Ma:2016mwh}, should be absent. This can be achieved by resorting to some softly broken symmetry or forcing the fermion-fermion-scalar loop vertex to contain a derivative on the fermion with correct chirality~\cite{Cepedello:2018rfh,CentellesChulia:2019xky}. In what follows, we will consider the softly broken non-Abelian $S_4$ symmetry and apply it to the diagram T1-i-a as a demonstration of how the absence of these lower-order contributions is guaranteed.   

The discrete flavor group $S_4$ has been used to predict the lepton-flavor mixing angles and the $CP$-violation phases~\cite{Ma:2005pd,Ding:2009iy,Hagedorn:2010th,Ishimori:2010au,Ding:2013hpa} and, recently, to ensure the dominance of the one-loop contributions to the Dirac neutrino masses~\cite{Yao:2017vtm,Yao:2018ekp}. It has five irreducible representations, two singlets $\mathbf{1}$ and $\mathbf{1}'$, one doublet $\mathbf{2}$, and two triplets $\mathbf{3}$ and $\mathbf{3}'$, with their tensor decomposition rules given, respectively, by~\cite{Kobayashi:2022moq} 
\begin{align} \label{eq:tensordecomposition}
& \mathbf{1}'\otimes \mathbf{1}'=\mathbf{1}, \quad \mathbf{1}'\otimes \mathbf{2}=\mathbf{2}, \quad  \mathbf{1}'\otimes \mathbf{3}=\mathbf{3}', 
\quad \mathbf{1}'\otimes \mathbf{3}'=\mathbf{3},  \nonumber \\[0.12cm]
&\mathbf{2}\otimes \mathbf{2}=\mathbf{1}\oplus \mathbf{1}'\oplus \mathbf{2}, \quad  
\mathbf{2}\otimes \mathbf{3}=\mathbf{2}\otimes \mathbf{3}'=\mathbf{3}\oplus \mathbf{3}', \nonumber \\[0.12cm]
&\mathbf{3}\otimes \mathbf{3}=\mathbf{3}'\otimes \mathbf{3}'=\mathbf{1}\oplus \mathbf{2}\oplus\mathbf{3}\oplus \mathbf{3}', \nonumber \\[0.12cm]
& \mathbf{3}\otimes \mathbf{3}'=\mathbf{1}'\oplus \mathbf{2}\oplus\mathbf{3}\oplus \mathbf{3}'.
\end{align}

We assign all the SM fermions to be $\mathbf{3}$, while the right-handed neutrinos $\nu_R$ to be $\mathbf{3}'$. Note that certain LQs, such as $S_1$, $\widetilde{R}_2$, $U_1$, and $\widetilde{V}_2$, have couplings with both the SM fermions and the right-handed neutrinos, and hence can generate the Dirac neutrino masses through the one-loop diagrams shown in Fig.~\ref{fig:dim4-oneloop}, as discussed in Ref.~\cite{Ma:2016mwh}. To exclude such a possibility, we divide the LQs into the following two groups: the first one refers to the LQs that couple with $\nu_R$ and transform as $\mathbf{1'}$, while the second one to the LQs transforming as $\mathbf{3}$ under the $S_4$ symmetry. Then, the one-loop diagrams shown in Fig.~\ref{fig:dim4-oneloop} are excluded due to the violation of the $S_4$ symmetry. In addition, as indicated in Table~\ref{tab:Symmetry}, the renormalizable Yukawa interaction $\bar{L}_L\widetilde{H}\nu_R$ and the seesaw realizations of the effective operator $\mathcal{L}_{4}$ are automatically forbidden by the $S_4$ symmetry, while 
the two-loop diagram T1-i-a can be generated due to the soft-breaking term $X^{\dagger}_5X_4X_3H$. Thus, the $S_4$ symmetry introduced guarantees the two-loop dominance in generating the Dirac neutrino masses. It should be, however, mentioned that the representation assignment here is solely as an illustration. It may not be the best choice, since, besides the Dirac neutrino masses, a compatible Pontecorvo-Maki-Nakagawa-Sakat mixing matrix must also be reproduced, once a specific UV model is concerned under the symmetry. It should also be pointed out that the choice of the auxiliary symmetry is highly subjective; other symmetries, either Abelian (e.g., $Z_{2}$, $Z_3$~\cite{Yao:2017vtm}) or non-Abelian (e.g., $A_4$~\cite{Yao:2018ekp}),  have also been considered to secure the dominance of the loop contributions.      

\begin{table}[t]
	\renewcommand\arraystretch{1.5} 
	\tabcolsep=0.26cm
	\centering
	\begin{tabular}{ccccccccc}
		\hline \hline
		& \multicolumn{8}{c}{T1-i-a}\\ \cline{2-9}
		Fields &  $L_L$ &$\nu_R$ & $H$ & $X_1$ & $X_2$ & $X_3$ & $X_4$  & $X_5$
		\\ \hline
		$Z_2$ & $+$ & $+$ & $+$ & $+$ & $-$ & $-$ & $-$ & $+$   \\
		$S_4$ & $\mathbf{3}$ & $\mathbf{3}'$ & $\mathbf{1}$ & $\mathbf{3}$ & $\mathbf{3}'$ & $\mathbf{3}$ & $\mathbf{1}'$  
		& $\mathbf{3}$  \\ 
		\hline \hline
	\end{tabular}
	\caption{Forbidding the lower-order contributions in the UV models of the diagram T1-i-a under the $S_4$ symmetry. We list also the transformation properties of the relevant fields under the $Z_2$ symmetry.}
	\label{tab:Symmetry} 
\end{table}

In short, once the dominance of the two-loop contributions to the Dirac neutrino masses is established, the UV completions listed in Tables~\ref{tab:T1-i-a}--\ref{tab:T3-x-a} can be identified as the LQ-SD$\nu$M, and the mighty ``stones'' we are seeking, if existed, must arise from them.
  
\section{The mighty ``stones''}
\label{sec:mighty_stones}

If the LQs proposed to account for the $R_{D^{(\ast)}}$, $R_{K^{(\ast)}}$, and $(g-2)_{\mu}$ anomalies also emerge in our established LQ-SD$\nu$M, there will be a good chance that these models are the mighty ``stones'' aimed at providing a simultaneous explanation of the flavor anomalies and the neutrino masses in a unified picture. To this end, we will firstly find out the proper LQs. 

\begin{table}[t]
	\renewcommand\arraystretch{1.5} 
	\tabcolsep=0.62cm
	\centering
	\begin{tabular}{cccc}
		\hline \hline
		Model &  $R_{K^{(\ast)}}$ & $R_{D^{(\ast)}}$ & $(g-2)_{\mu}$\\ \hline
		$U_1$ & \color{red}\cmark & \color{red}\cmark & \color{red}\cmark \\ 
		$V_2$ & \color{red}\cmark & \color{red}\cmark & \color{red}\cmark \\ 
		$S_1$ & \color{blue}\xmark & \color{red}\cmark & \color{red}\cmark \\ 
		$S_3$ & \color{red}\cmark & \color{blue}\xmark & \color{blue}\xmark \\ 
		$R_2$ & \color{blue}\xmark & \color{red}\cmark & \color{red}\cmark \\  \hline \hline
	\end{tabular}
	\caption{Summary of the LQs that can accommodate the $R_{K^{(\ast)}}$, $R_{D^{(\ast)}}$, and/or $(g-2)_{\mu}$ anomalies.}
	\label{tab:anomalies}    
\end{table}

It is known that the vector LQ $U_1$ can alone address the $R_{D^{(\ast)}}$ and $R_{K^{(\ast)}}$ anomalies simultaneously, while the LQs $S_{1}$, $S_{3}$, $R_{2}$, $\widetilde{R}_{2}$, and $U_3$ can only account for one of the anomalies~\cite{Angelescu:2018tyl,Angelescu:2021lln}. Interestingly, it has been shown recently in Refs.~\cite{Du:2021zkq,Ban:2021tos} that the same vector LQ $U_1$ can also explain the $(g-2)_{\mu}$ anomaly~\cite{Muong-2:2006rrc,Muong-2:2021ojo,Aoyama:2020ynm}, provided that its interactions with the left- and right-handed SM fermions are both present. These findings lead to an exciting observation that all the flavor anomalies can be addressed by such a single vector LQ. Remarkably, $U_1$ is in fact not alone. A very recent study shows that the same task can also be achieved by another vector LQ $V_2$~\cite{Cheung:2022zsb}. 

Besides the option with a single vector LQ $U_1$ or $V_2$, one could also resort to a pair of scalar LQs. It is known that at tree-level $S_3$ is the only scalar LQ that can account for the $R_{K^{(\ast)}}$ anomalies, while either $S_1$ or $R_2$ can address the $R_{D^{(\ast)}}$ anomalies (see, e.g., Refs.~\cite{Li:2016vvp,Buttazzo:2017ixm,Angelescu:2019eoh,Angelescu:2021lln} and references therein). Meanwhile, both $S_1$ and $R_2$ can address the $(g-2)_{\mu}$ anomaly, due to the simultaneous presence of their couplings to the SM left- and right-handed chiral fermions~\cite{Cheung:2001ip}. Thus, to solve all the flavor anomalies, the first step is to select the proper LQ pairs. In order to help visualize these possible options, we summarize in Table~\ref{tab:anomalies} the aforementioned LQs that can accommodate the $R_{K^{(\ast)}}$, $R_{D^{(\ast)}}$, and/or $(g-2)_{\mu}$ anomalies. It becomes clear that, besides the vector LQs $U_1$ and $V_2$, we have two more options with a pair of LQs: (i) $S_1$ and $S_3$, and (ii) $R_2$ and $S_3$. Both options have been explored to address these flavor anomalies in a unified framework~\cite{Chen:2017hir,Bigaran:2019bqv,Crivellin:2019dwb,Saad:2020ihm,Babu:2020hun,Lee:2021jdr,Julio:2022bue,DAlise:2022ypp}.

We are now ready to establish the mighty ``stones''. Given the high freedom in choosing the auxiliary 
symmetry to secure the two-loop dominance, we will directly match the LQs in Table~\ref{tab:anomalies} with those in the UV models listed in Tables~\ref{tab:T1-i-a}--\ref{tab:T3-x-a}. For the options $U_1$ and $V_2$, clearly plenty of models are available. Among them, the ones associated with the diagrams T1-i-a and T3-iii-a are probably the optimum for both options, since a minimum of new fields are involved. Here it should be mentioned that the models associated with the diagrams T1-i-b, T3-iv-b, T3-ix-a, and T3-x-a can be optimal for the option $U_1$ as well, due to the same reason.  

However, for the scalar-LQ options (i) and (ii), none of the models listed in Tables~\ref{tab:T1-i-a}--\ref{tab:T3-x-a} are available, because at least an additional LQ has to be introduced. Under such a circumstance, for the scalar-LQ option (i), the model A-1 (A-2) associated with the diagrams T1-i-a, T1-i-b,T3-iii-a, T3-iv-b, T3-ix-a, and T3-x-a, after extended to include the missing LQ $S_1$ or $S_3$, shall be the optimum, because it will involve a minimal number of new degrees of freedom. Intriguingly, both $S_1$ and $S_3$ will contribute to the neutrino mass generation in the enlarged model A-1 (A-2) of the diagrams T1-i-a and T3-iii-a. The next best option would be the model A-1 (A-2) associated, e.g., with the diagrams T3-i-a, T3-i-b, T3-ii-a, T3-ii-b, T2-i-a, T2-i-b, T3-iii-b, and T3-iv-a, since only one additional new field must be added. Similar to the previous optimum, both $S_1$ and $S_3$ in the modified model A-1 (A-2) of the diagrams T3-i-a, T3-ii-a, T2-i-a, and T3-iv-a will contribute to the neutrino mass generation; interestingly, the updated model T3-ii-a-A-1 (A-2) will involve overall three scalar LQs. 

Semi-similar conclusions drawn for the scalar-LQ option (i) hold for the option (ii) as well. For instance, the optimum shall arise from the model B-1 (B-2) associated with the diagrams T1-i-a and T3-iii-a enlarged with the LQ $S_3$, while the next best option from the updated model B-1 (B-2) associated with the diagrams T3-i-a, T3-ii-a, T2-i-a, and T3-iv-a. 

At this point, it may be interesting to note that combining two of the above models---each contains one component of the scalar LQ pair---surely works for both the options (i) and (ii). Although these combined models often involve more new degrees of freedom and are therefore less appealing, some of them can still be very interesting. For instance, the models A-1 (A-2) and B-1 (B-2) associated with the diagram T1-i-a, although each of them consists of three new fields besides the LQs, share two of them. Bringing the two models together can thus generate another best option as well, which consists of six new degrees of freedom, just like the aforementioned updated models such as T3-i-a-A-1 (A-2). Besides, both the LQs $S_3$ and $R_2$ in this model will contribute to the neutrino mass generation. Based on this model, one may also build an even larger model by adding the LQ $S_1$. Although the model seems cumbersome, a slight compromise on minimality might be a good bargain here, given especially that it contains all the scalar LQs in Table~\ref{tab:anomalies}.

\begin{figure}[t]
	\centering
	\includegraphics[width=0.29\textwidth]{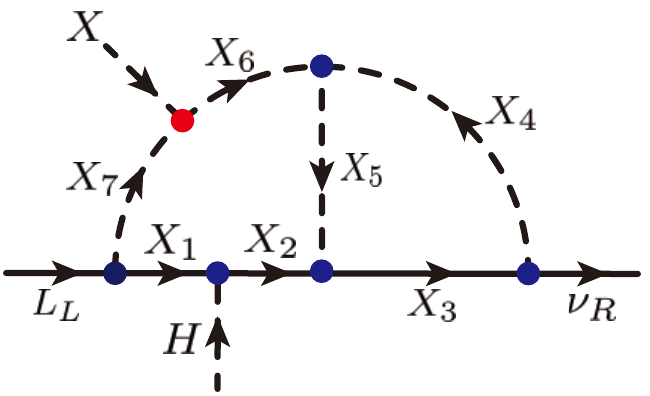}
	\caption{Generating a two-loop realization of $\mathcal{L}_5$ by linking $X$ to $X_6$ in the diagram T3-iii-a depicted in Fig.~\ref{fig:4d2loop}.}
	\label{fig:5d} 
\end{figure}

Finally, we conclude this section by making the following comment. In the mighty models involving the vector LQ $U_1$ or $V_2$, there is a close-knit connection between the flavor anomalies and the neutrino masses. In other words, the absence of any particle introduced to account for the flavor anomalies in these mighty models will fail to generate the neutrino masses.\footnote{The same idea has been appreciated recently in Refs.~\cite{Babu:2020hun,Julio:2022bue}.} In the mighty models consisting of the scalar LQs listed in Table~\ref{tab:anomalies}, on the other hand, such a connection is missing, since removing $S_1$ or $S_3$ from the enlarged model A-1 (A-2) associated with, e.g., the diagram T1-i-a does not render the neutrino massless. To build a close-knit connection for the scalar-LQ options (i) and (ii), one can consider, e.g., the two-loop realizations of the effective operator $\mathcal{L}_5=-g \bar{\nu}_RL_L HX+\text{H.c.}$, where $X$ is a color-singlet, $\text{SU}(2)_{\text{L}}$-triplet scalar with hypercharge $\alpha=0$. Given that the two-loop diagrams of $\mathcal{L}_5$ and $\mathcal{L}_4$ share similar structures (or skeletons)~\cite{CentellesChulia:2019xky,Jana:2019mgj}, one can obtain the former (at least some of them) by attaching the field $X$ to the propagators of the latter, and in turn build the corresponding UV models. Let us take the model T3-iii-a-A-1 as an illustration. We firstly set $X_3=(1,1,\alpha)$, $X_4=(1,1,\alpha)$, $X_5=(3,1,\alpha\!-\!1/3)$, and $X_6=S_{1}$ or $S_{3}$. Then, as shown in Fig.~\ref{fig:5d}, we can obtain a two-loop realization of $\mathcal{L}_5$ by linking $X$ to $X_6$, and build the corresponding UV model, which contains, besides the fields $X_{3,4,5}$, both $S_1$ and $S_3$. We can further impose an auxiliary symmetry to forbid all the lower-order diagrams, as well as the one T3-iii-a, while introducing a soft-breaking interaction term (marked in red in Fig.~\ref{fig:5d}) to support the two-loop dominance. In this way, both $S_1$ and $S_3$ become indispensable for the neutrino mass generation, and thus a close-knit connection is established between the flavor anomalies and the neutrino masses. 
 
\section{Conclusion}
\label{sec:con}

In this paper, we have pointed out a potential pathway to address in a unified picture the flavor anomalies and the origin of neutrino masses. The key ingredient here rests on the compatibility of the LQs proposed to account for the $R_{D^{(\ast)}}$, $R_{K^{(\ast)}}$, and $(g-2)_\mu$ anomalies with the SD$\nu$M, which is featured by its capability to explain the neutrino mass generation and the DM property. Based on the minimal seesaw, one-loop, 
and two-loop realizations of the effective operators $\mathcal{L}_{4}$ for the Dirac neutrino masses, and guided by the topology-selection criteria outlined in Sec.~\ref{sec:classfication}, we have found that plenty of diagrams in the two-loop realizations of $\mathcal{L}_4$ can support the coexistence of the LQs and the DM candidates, and exhausted in Tables~\ref{tab:T1-i-a}--\ref{tab:T3-x-a} all the topology-based UV completions. Matching the LQs of these UV models with the ones introduced to accommodate all the flavor anomalies considered, we have established the mighty models that can address all the aforementioned problems in a unified picture. On top of that, we have also found that, in the models involving the vector LQ $U_1$ or $V_2$, a close-knit connection can be established between the flavor anomalies and the neutrino masses.
    
There exists, without any doubt, very rich phenomenology for each mighty model. However, since  
a detailed analysis has to be done on a case-by-case basis, we will leave it for future works.

\section*{Acknowledgments}
We are grateful to Ricardo Cepedello for helpful discussions. This work is supported by the National Natural Science Foundation of China under Grant Nos.~12135006, 12075097, 12047527, 11675061, and 11775092, as well as by the Fundamental Research Funds for the Central Universities under Grant Nos.~CCNU20TS007 and CCNU22LJ004.

\appendix
\renewcommand{\theequation}{A.\arabic{equation}}

\section{\boldmath Systematic classification of tree-level and one-loop realizations of the operator $\mathcal{L}_{4}$}
\label{sec:appA}

In this appendix, following Refs.~\cite{Ma:2016mwh,Yao:2018ekp}, we will present a systematic classification of tree-level and one-loop realizations of the effective operator $\mathcal{L}_{4}$. 

\begin{figure}[t]
	\centering
	\includegraphics[width=0.48\textwidth]{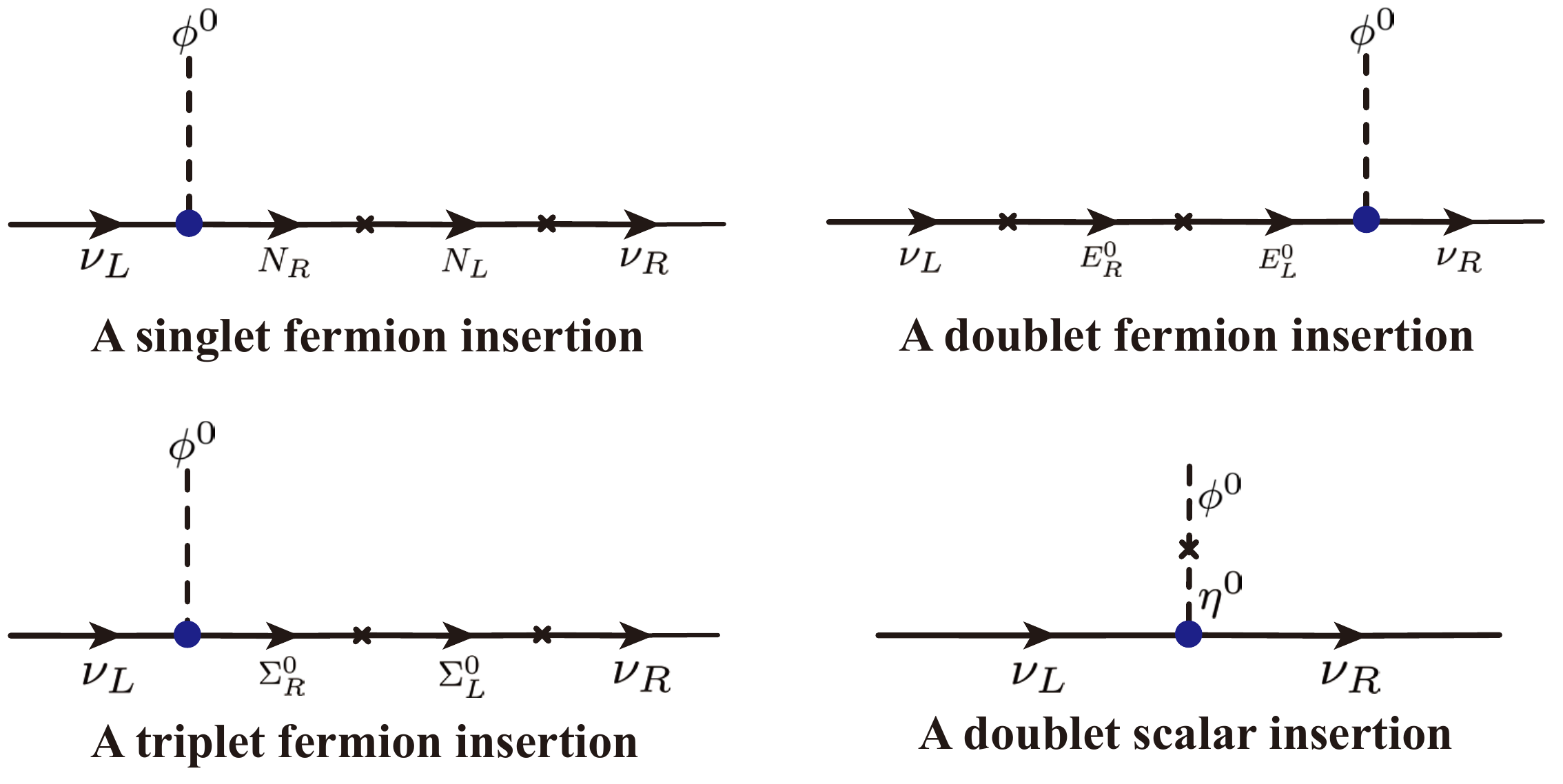}
	\caption{Four different seesaw realizations of the Dirac neutrino mass operator $\mathcal{L}_{4}$ with different new particle insertions, where $N$, $E=(E^0,E^-)^T$, and $\Sigma=(\Sigma^+,\Sigma^0,\Sigma^-)^T$ denote a Dirac singlet, a Dirac doublet, and a Dirac triplet fermion respectively, whereas $\eta=(\eta^+,\eta^0)^T$ is a scalar doublet~\cite{Ma:2016mwh}. }
	\label{fig:dim4-seesaw} 
\end{figure}

The tree-level and one-loop realizations of the Dirac neutrino mass operator $\mathcal{L}_{4}$ have been studied in Ref.~\cite{Ma:2016mwh}. It is shown that there are only four ways to achieve the seesaw (tree-level) realizations, i.e., the Dirac neutrino masses can be generated with the insertion of a Dirac singlet $N$, a Dirac doublet $E=(E^0,E^-)^T$, a Dirac triplet fermion $\Sigma=(\Sigma^+,\Sigma^0,\Sigma^-)^T$, or a doublet scalar $\eta=(\eta^+,\eta^0)^T$. The corresponding Feynman diagrams are depicted in Fig.~\ref{fig:dim4-seesaw}.

\begin{figure}[t]
	\centering
	\includegraphics[width=0.48\textwidth]{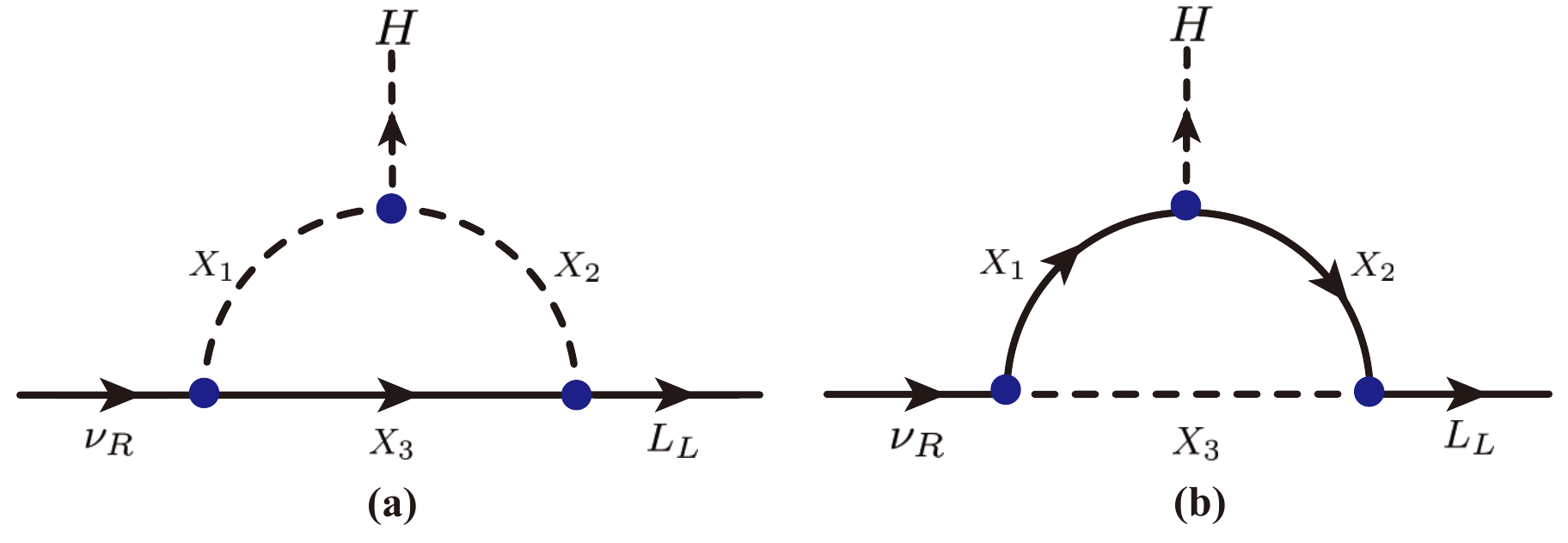}
	\caption{One-loop realizations of the Dirac neutrino mass operator $\mathcal{L}_{4}$. The dashed line represents either a scalar or a vector field, while the solid line a fermion one.}
	\label{fig:dim4-oneloop}
\end{figure}

If the seesaw realizations discussed above are not available, the Dirac neutrino masses may still arise at the one-loop level. As demonstrated in Ref.~\cite{Ma:2016mwh}, there are only two Feynman diagrams for the one-loop generation of the Dirac neutrino masses, which are shown in Fig.~\ref{fig:dim4-oneloop}. It can be seen that both diagrams contain three propagators, two bosonic and one fermionic in Fig.~\ref{fig:dim4-oneloop}\,(a), whereas one bosonic and two fermionic in Fig.~\ref{fig:dim4-oneloop}\,(b). With the appropriate assignments of the SM gauge charges to the fields $X_{1,2,3}$, the DM candidates can potentially arise from them or their neutral components~\cite{Ma:2016mwh,Yao:2017vtm}. If so, the corresponding UV models can then be identified as the SD$\nu$M. 

\section{UV completions and DM candidates}
\label{sec:appB}
 
In Sec.~\ref{sec:classfication}, we have shown that the diagrams depicted in Fig.~\ref{fig:4d2loop} support the coexistence of the LQs and the DM candidates. To pinpoint the possible quantum numbers of the messenger fields, we have then explored possible assignments of the $Z_2$ parity and the color $c$ to these fields, and found that the combinations ``a'' and ``b'' can support our selection criteria better. In this appendix, guided by the two combinations and our selection criteria, we will present in Tables~\ref{tab:T1-i-a}--\ref{tab:T3-x-a} the possible quantum numbers of all the internal fields in each diagram and, at the same time, the UV models resulting from the diagram. Although we have uniformly labeled the models in these tables by x-y with x=A, B, C, $\cdots$ and y=1, 2, 3, $\cdots$, they are properly referred by, e.g., T1-i-a-x-y in the main text, where T1-i-a denotes the diagram T1-i with the combination ``a''. To help identify the possible DM candidates, we have specified the fermionic messenger fields and their transformation properties under the $Z_2$ symmetry. By setting down their hypercharges according to our prescriptions for the DM candidates, we have also selected them in each model. 

\begin{table*}[htbp] 
	\renewcommand\arraystretch{1.5}  
	\centering 
	\tabcolsep=0.57cm
	\begin{tabular}{ccccccc}
		\hline\hline
		Model & $X^F_1$ & $X^F_2$ & $X_3$ & $X_4$  & $X_5$ & DM \\ \hline
		A-1     & $\bar{Q}_L$ & $(1,1\oplus3,\alpha)^-$ & $(1,1\oplus3,\alpha)^-$ & $(3,2,\alpha\!+\!\frac{1}{6})^-$ & $S_{1,3}^{\dagger}$ & $[X_2,X_3]|_{\alpha=0}$\\
		A-2     & $\bar{Q}_L$ & $(1,2,\alpha)^-$ & $(1,2,\alpha)^-$ & $(3,1\oplus3,\alpha\!+\!\frac{1}{6})^-$ & $S_{1,3}^{\dagger}$ & $X_3|_{\alpha=\frac{1}{2}}$\\ 
		B-1     & $u_R$ & $(1,1\oplus3,\alpha)^-$ & $(1,1\oplus3,\alpha)^-$ & $(\bar{3},1,\alpha\!-\!\frac{2}{3})^-$ & $R_2^{\dagger}$ & $[X_2,X_3]|_{\alpha=0}$\\
		B-2     & $u_R$  & $(1,2,\alpha)^-$ & $(1,2,\alpha)^-$ & $(\bar{3},2,\alpha\!-\!\frac{2}{3})^-$ & $R_2^{\dagger}$ & $X_3|_{\alpha=\frac{1}{2}}$\\
		C-1     & $d_R$ & $(1,1\oplus3,\alpha)^-$ & $(1,1\oplus3,\alpha)^-$ & $(\bar{3},1\oplus3,\alpha\!+\!\frac{1}{3})^-$ & $\widetilde{R}_2^{\dagger}$ & $[X_2,X_3]|_{\alpha=0}$\\
		C-2     & $d_R$  & $(1,2,\alpha)^-$ & $(1,2,\alpha)^-$ & $(\bar{3},2,\alpha\!+\!\frac{1}{3})^-$ & $\widetilde{R}_2^{\dagger}$ & $X_3|_{\alpha=\frac{1}{2}}$\\
		D-1     & $Q_L$ & $(1,1\oplus3,\alpha)^-$ & $(1,1\oplus3,\alpha)^-$ & $(\bar{3},2,\alpha\!-\!\frac{1}{6})^-$ & $U_{1,3}^{\dagger}$ & $[X_2,X_3]|_{\alpha=0}$\\
		D-2     & $Q_L$ & $(1,2,\alpha)^-$ & $(1,2,\alpha)^-$ & $(\bar{3},1\oplus3,\alpha\!-\!\frac{1}{6})^-$ & $U_{1,3}^{\dagger}$ & $X_3|_{\alpha=\frac{1}{2}}$ \\
		E-1     & $\bar{u}_R$ & $(1,1\oplus3,\alpha)^-$ & $(1,1\oplus3,\alpha)^-$ & $(3,1\oplus3,\alpha\!+\!\frac{2}{3})^-$ & $\widetilde{V}_2^{\dagger}$ & $[X_2,X_3]|_{\alpha=0}$\\
		E-2     & $\bar{u}_R$  & $(1,2,\alpha)^-$ & $(1,2,\alpha)^-$ & $(3,2,\alpha\!+\!\frac{2}{3})^-$ & $\widetilde{V}_2^{\dagger}$ & $X_3|_{\alpha=\frac{1}{2}}$\\
		F-1     & $\bar{d}_R$ & $(1,1\oplus3,\alpha)^-$ & $(1,1\oplus3,\alpha)^-$ & $(3,1\oplus3,\alpha\!-\!\frac{1}{3})^-$ & $V_2^{\dagger}$ & $[X_2,X_3]|_{\alpha=0}$\\
		F-2     & $\bar{d}_R$  & $(1,2,\alpha)^-$ & $(1,2,\alpha)^-$ & $(3,2,\alpha\!-\!\frac{1}{3})^-$ & $V_2^{\dagger}$ & $X_3|_{\alpha=\frac{1}{2}}$\\
		\hline\hline		
	\end{tabular}
	\caption{Possible assignments of the quantum numbers under the $\text{SU(3)}_{c}\otimes \text{SU}(2)_{\text{L}}\otimes \text{U}(1)_{\alpha}\otimes Z_2$ symmetry to the mediators in the diagram T1-i depicted in Fig.~\ref{fig:4d2loop} and with the combination (a)-(A) shown in Fig.~\ref{fig:Z2}, where $m\oplus n$ implies that either an SU(2)$_{\text{L}}$ $m$- or $n$-plet works. The possible DM candidates are identified by the hypercharge conditions for the fields, as indicated by the last column, where the symbols like $[X_2,X_3]|_{\alpha=0}$ require both the fields $X_2$ and $X_3$ to have the same hypercharge or hypercharge component $\alpha=0$.}
	\label{tab:T1-i-a}   
\end{table*}

\begin{table*}[htbp] 
	\renewcommand\arraystretch{1.5}  
	\centering 
	\tabcolsep=0.512cm
	\begin{tabular}{ccccccc}
		\hline\hline
		Model & $X^F_1$ & $X^F_2$ & $X_3$ & $X_4$  & $X_5$ & DM \\ \hline
		A-1     & $(1,1\oplus3,\alpha)^-$  & $\bar{d}_R$ & $S_{1}$ & $(\bar{3},1\oplus3,\frac{1}{3}\!-\!\alpha)^-$ & $(1,2,-\frac{1}{2}\!-\!\alpha)^-$ & $[X_1,X_5]|_{\alpha=0}$\\
		A-2     & $(1,2,\alpha)^-$  & $\bar{d}_R$ & $S_{1}$ & $(\bar{3},2,\frac{1}{3}\!-\!\alpha)^-$ & $(1,1\oplus3,-\frac{1}{2}\!-\!\alpha)^-$ & $X_5|_{\alpha=-\frac{1}{2}}$\\ 
		B-1     & $(1,1\oplus3,\alpha)^-$  & $Q_L$ & $\widetilde{R}_{2}$ & $(3,2,\frac{1}{6}\!-\!\alpha)^-$ & $(1,2,-\frac{1}{2}\!-\!\alpha)^-$ & $[X_1,X_5]|_{\alpha=0}$\\
		B-2     & $(1,2,\alpha)^-$  & $Q_L$ & $\widetilde{R}_{2}$ & $(3,1\oplus3,\frac{1}{6}\!-\!\alpha)^-$ & $(1,1\oplus3,-\frac{1}{2}\!-\!\alpha)^-$ & $X_5|_{\alpha=-\frac{1}{2}}$\\ 
		C-1     & $(1,1\oplus3,\alpha)^-$  & $\bar{u}_R$ & $\bar{S}_{1}$ & $(\bar{3},1\oplus3,-\frac{2}{3}\!-\!\alpha)^-$ & $(1,2,-\frac{1}{2}\!-\!\alpha)^-$ & $[X_1,X_5]|_{\alpha=0}$\\
		C-2     & $(1,2,\alpha)^-$  & $\bar{u}_R$ & $\bar{S}_{1}$ & $(\bar{3},2,-\frac{2}{3}\!-\!\alpha)^-$ & $(1,1\oplus3,-\frac{1}{2}\!-\!\alpha)^-$ & $X_5|_{\alpha=-\frac{1}{2}}$\\ 
		D-1     & $(1,1\oplus3,\alpha)^-$  & $u_R$ & $U_{1}$ & $(3,1\oplus3,\frac{2}{3}\!-\!\alpha)^-$ & $(1,2,-\frac{1}{2}\!-\!\alpha)^-$ & $[X_1,X_5]|_{\alpha=0}$\\
		D-2     & $(1,2,\alpha)^-$  & $u_R$ & $U_{1}$ & $(3,2,\frac{2}{3}\!-\!\alpha)^-$ & $(1,1\oplus3,-\frac{1}{2}\!-\!\alpha)^-$ & $X_5|_{\alpha=-\frac{1}{2}}$\\ 
		E-1     & $(1,1\oplus3,\alpha)^-$  & $\bar{Q}_L$ & $\widetilde{V}_{2}$ & $(\bar{3},2,-\frac{1}{6}\!-\!\alpha)^-$ & $(1,2,-\frac{1}{2}\!-\!\alpha)^-$ & $[X_1,X_5]|_{\alpha=0}$\\
		E-2     & $(1,2,\alpha)^-$  & $\bar{Q}_L$ & $\widetilde{V}_{2}$ & $(\bar{3},1\oplus3,-\frac{1}{6}\!-\!\alpha)^-$ & $(1,1\oplus3,-\frac{1}{2}\!-\!\alpha)^-$ &$X_5|_{\alpha=-\frac{1}{2}}$\\ 
		F-1     & $(1,1\oplus3,\alpha)^-$  & $d_R$ & $\bar{U}_{1}$ & $(3,1\oplus3,-\frac{1}{3}\!-\!\alpha)^-$ & $(1,2,-\frac{1}{2}\!-\!\alpha)^-$ & $[X_1,X_5]|_{\alpha=0}$\\
		F-2     & $(1,2,\alpha)^-$  & $d_R$ & $\bar{U}_{1}$ & $(3,2,-\frac{1}{3}\!-\!\alpha)^-$ & $(1,1\oplus3,-\frac{1}{2}\!-\!\alpha)^-$ & $X_5|_{\alpha=-\frac{1}{2}}$\\ 
		\hline\hline		
	\end{tabular}
	\caption{Same as in Table~\ref{tab:T1-i-a} but for the diagram T1-i depicted in Fig.~\ref{fig:4d2loop} and with the combination (b)-(B) shown in Fig.~\ref{fig:Z2}.}
	\label{tab:T1-i-b}   
\end{table*}

\begin{table*}[htbp] 
	\renewcommand\arraystretch{1.5}  
	\centering 
	\tabcolsep=0.295cm
	\begin{tabular}{cccccccc}
		\hline\hline
		Model & $X^F_1$ & $X^F_2$ & $X_3$ & $X_4$  & $X_5$ & $X_6$ & DM \\ \hline
		A-1     & $\bar{Q}_L$ & $(1,1\oplus3,\alpha)^-$ & $(1,1\oplus3,\alpha)^-$ &$(1,2,\alpha+\frac{1}{2})^-$ & $(3,2,\alpha\!+\!\frac{1}{6})^-$ & $S_{1,3}^{\dagger}$ & $[X_2,X_3,X_4]|_{\alpha=0}$\\
		A-2     & $\bar{Q}_L$ & $(1,2,\alpha)^-$ & $(1,2,\alpha)^-$ &$(1,1\oplus3,\alpha+\frac{1}{2})^-$ & $(3,1\oplus3,\alpha\!+\!\frac{1}{6})^-$ & $S_{1,3}^{\dagger}$ & $[X_3,X_4]|_{\alpha=-\frac{1}{2}}$\\
		B-1     & $u_R$ & $(1,1\oplus3,\alpha)^-$ & $(1,1\oplus3,\alpha)^-$ &$(1,2,\alpha+\frac{1}{2})^-$ & $(\bar{3},1,\alpha\!-\!\frac{2}{3})^-$ & $R_2^{\dagger}$ & $[X_2,X_3,X_4]|_{\alpha=0}$\\
		B-2     & $u_R$  & $(1,2,\alpha)^-$ & $(1,2,\alpha)^-$& $(1,1\oplus3,\alpha+\frac{1}{2})^-$ & $(\bar{3},2,\alpha\!-\!\frac{2}{3})^-$ & $R_2^{\dagger}$ & $[X_3,X_4]|_{\alpha=-\frac{1}{2}}$\\
		C-1     & $d_R$ & $(1,1\oplus3,\alpha)^-$ & $(1,1\oplus3,\alpha)^-$ &$(1,2,\alpha+\frac{1}{2})^-$  & $(\bar{3},1\oplus3,\alpha\!+\!\frac{1}{3})^-$ & $\widetilde{R}_2^{\dagger}$ & $[X_2,X_3,X_4]|_{\alpha=0}$\\
		C-2     & $d_R$  & $(1,2,\alpha)^-$ & $(1,2,\alpha)^-$& $(1,1\oplus3,\alpha+\frac{1}{2})^-$  & $(\bar{3},2,\alpha\!+\!\frac{1}{3})^-$ & $\widetilde{R}_2^{\dagger}$ & $[X_3,X_4]|_{\alpha=-\frac{1}{2}}$ \\
		D-1     & $Q_L$ & $(1,1\oplus3,\alpha)^-$ & $(1,1\oplus3,\alpha)^-$ &$(1,2,\alpha+\frac{1}{2})^-$  & $(\bar{3},2,\alpha\!-\!\frac{1}{6})^-$ & $U_{1,3}^{\dagger}$ & $[X_2,X_3,X_4]|_{\alpha=0}$\\
		D-2     & $Q_L$ & $(1,2,\alpha)^-$ & $(1,2,\alpha)^-$ & $(1,1\oplus3,\alpha+\frac{1}{2})^-$ & $(\bar{3},1\oplus3,\alpha\!-\!\frac{1}{6})^-$ & $U_{1,3}^{\dagger}$ &  $[X_3,X_4]|_{\alpha=-\frac{1}{2}}$ \\
		E-1     & $\bar{u}_R$ & $(1,1\oplus3,\alpha)^-$ & $(1,1\oplus3,\alpha)^-$&$(1,2,\alpha+\frac{1}{2})^-$  & $(3,1\oplus3,\alpha\!+\!\frac{2}{3})^-$ & $\widetilde{V}_2^{\dagger}$ & $[X_2,X_3,X_4]|_{\alpha=0}$\\
		E-2     & $\bar{u}_R$  & $(1,2,\alpha)^-$ & $(1,2,\alpha)^-$ & $(1,1\oplus3,\alpha+\frac{1}{2})^-$ &$(3,2,\alpha\!+\!\frac{2}{3})^-$ & $\widetilde{V}_2^{\dagger}$ & $[X_3,X_4]|_{\alpha=-\frac{1}{2}}$ \\
		F-1     & $\bar{d}_R$ & $(1,1\oplus3,\alpha)^-$ & $(1,1\oplus3,\alpha)^-$&$(1,2,\alpha+\frac{1}{2})^-$  & $(3,1\oplus3,\alpha\!-\!\frac{1}{3})^-$ & $V_2^{\dagger}$ & $[X_2,X_3,X_4]|_{\alpha=0}$\\
		F-2     & $\bar{d}_R$  & $(1,2,\alpha)^-$ & $(1,2,\alpha)^-$ & $(1,1\oplus3,\alpha+\frac{1}{2})^-$ & $(3,2,\alpha\!-\!\frac{1}{3})^-$ & $V_2^{\dagger}$ & $[X_3,X_4]|_{\alpha=-\frac{1}{2}}$ \\
		\hline\hline		
	\end{tabular}
	\caption{Same as in Table~\ref{tab:T1-i-a} but for the diagram T3-i depicted in Fig.~\ref{fig:4d2loop} and with the combination (a)-(A) shown in Fig.~\ref{fig:Z2}.}
	\label{tab:T3-i-a}   
\end{table*}

\begin{table*}[htbp] 
	\renewcommand\arraystretch{1.5}  
	\centering 
	\tabcolsep=0.365cm
	\begin{tabular}{cccccccc}
		\hline\hline
		Model & $X^F_1$ & $X^F_2$ & $X_3$ & $X_4$  & $X_5$ & $X_6$ & DM \\ \hline
		A-1     & $(1,1\oplus3,\alpha)^-$  & $\bar{d}_R$ & $S_{1}$ & $(\bar{3},2,\frac{5}{6})$ & $(\bar{3},1\oplus3,\frac{1}{3}\!-\!\alpha)^-$ & $(1,2,-\frac{1}{2}\!-\!\alpha)^-$ & $[X_1,X_6]|_{\alpha=0}$\\
		A-2     & $(1,2,\alpha)^-$  & $\bar{d}_R$ & $S_{1}$ &  $(\bar{3},2,\frac{5}{6})$ & 
		$(\bar{3},2,\frac{1}{3}\!-\!\alpha)^-$ & $(1,1\oplus3,-\frac{1}{2}\!-\!\alpha)^-$ &$X_6|_{\alpha=-\frac{1}{2}}$\\ 
		B-1     & $(1,1\oplus3,\alpha)^-$  & $Q_L$ & $\widetilde{R}_{2}$ & $\bar{S}^{\dagger}_1$ 
		& $(3,2,\frac{1}{6}\!-\!\alpha)^-$& $(1,2,-\frac{1}{2}\!-\!\alpha)^-$ & $[X_1,X_6]|_{\alpha=0}$\\
		B-2     & $(1,2,\alpha)^-$  & $Q_L$ & $\widetilde{R}_{2}$& $\bar{S}^{\dagger}_1$
		& $(3,1\oplus3,\frac{1}{6}\!-\!\alpha)^-$ & $(1,1\oplus3,-\frac{1}{2}\!-\!\alpha)^-$ & $X_6|_{\alpha=-\frac{1}{2}}$\\ 
		C-1     & $(1,1\oplus3,\alpha)^-$  & $\bar{u}_R$ & $\bar{S}_{1}$ & $\widetilde{R}_2^{\dagger}$ &
		$(\bar{3},1\oplus3,-\frac{2}{3}\!-\!\alpha)^-$ & $(1,2,-\frac{1}{2}\!-\!\alpha)^-$ & $[X_1,X_6]|_{\alpha=0}$\\
		C-2     & $(1,2,\alpha)^-$  & $\bar{u}_R$ & $\bar{S}_{1}$ & $\widetilde{R}_2^{\dagger}$ &
		$(\bar{3},2,-\frac{2}{3}\!-\!\alpha)^-$ & $(1,1\oplus3,-\frac{1}{2}\!-\!\alpha)^-$ & $X_6|_{\alpha=-\frac{1}{2}}$\\
		D-1     & $(1,1\oplus3,\alpha)^-$  & $u_R$ & $U_{1}$ & $(3,2,\frac{7}{6})$ & $(3,1\oplus3,\frac{2}{3}\!-\!\alpha)^-$ & $(1,2,-\frac{1}{2}\!-\!\alpha)^-$ & $[X_1,X_6]|_{\alpha=0}$\\
		D-2     & $(1,2,\alpha)^-$  & $u_R$ & $U_{1}$ & $(3,2,\frac{7}{6})$&$(3,2,\frac{2}{3}\!-\!\alpha)^-$ & $(1,1\oplus3,-\frac{1}{2}\!-\!\alpha)^-$ & $X_6|_{\alpha=-\frac{1}{2}}$\\
		E-1     & $(1,1\oplus3,\alpha)^-$  & $\bar{Q}_L$ & $\widetilde{V}_{2}$ &$\bar{U}_1^{\dagger}$ & $(\bar{3},2,-\frac{1}{6}\!-\!\alpha)^-$ & $(1,2,-\frac{1}{2}\!-\!\alpha)^-$ & $[X_1,X_6]|_{\alpha=0}$\\
		E-2     & $(1,2,\alpha)^-$  & $\bar{Q}_L$ & $\widetilde{V}_{2}$ & $\bar{U}_1^{\dagger}$ & $(\bar{3},1\oplus3,-\frac{1}{6}\!-\!\alpha)^-$ & $(1,1\oplus3,-\frac{1}{2}\!-\!\alpha)^-$ & $X_6|_{\alpha=-\frac{1}{2}}$\\
		F-1     & $(1,1\oplus3,\alpha)^-$  & $d_R$ & $\bar{U}_{1}$ & $\widetilde{V}_2^{\dagger}$ & $(3,1\oplus3,-\frac{1}{3}\!-\!\alpha)^-$ & $(1,2,-\frac{1}{2}\!-\!\alpha)^-$ & $[X_1,X_6]|_{\alpha=0}$\\
		F-2     & $(1,2,\alpha)^-$  & $d_R$ & $\bar{U}_{1}$ & $\widetilde{V}_2^{\dagger}$  & $(3,2,-\frac{1}{3}\!-\!\alpha)^-$ & $(1,1\oplus3,-\frac{1}{2}\!-\!\alpha)^-$ & $X_6|_{\alpha=-\frac{1}{2}}$\\
		\hline\hline		
	\end{tabular}
	\caption{Same as in Table~\ref{tab:T1-i-a} but for the diagram T3-i depicted in Fig.~\ref{fig:4d2loop} and with the combination (b)-(B) shown in Fig.~\ref{fig:Z2}.}
	\label{tab:T3-i-b}   
\end{table*}

\begin{table*}[htbp] 
	\renewcommand\arraystretch{1.5}  
	\centering 
	\tabcolsep=0.445cm
	\begin{tabular}{cccccccc}
		\hline\hline
		Model & $X^F_1$ & $X^F_2$ & $X_3$ & $X_4$  & $X_5$ & $X_6$ & DM \\ \hline
		A-1     & $\bar{Q}_L$ & $(1,1\oplus3,\alpha)^-$ & $(1,1\oplus3,\alpha)^-$ & $(3,2,\alpha\!+\!\frac{1}{6})^-$ 
		& $\widetilde{R}_2$ &$S_{1,3}^{\dagger}$ & $[X_2,X_3]|_{\alpha=0}$\\
		A-2     & $\bar{Q}_L$ & $(1,2,\alpha)^-$ & $(1,2,\alpha)^-$ & $(3,1\oplus3,\alpha\!+\!\frac{1}{6})^-$ 
		&  $\widetilde{R}_2$  & $S_{1,3}^{\dagger}$ & $X_3|_{\alpha=\frac{1}{2}}$\\ 
		B-1     & $u_R$ & $(1,1\oplus3,\alpha)^-$ & $(1,1\oplus3,\alpha)^-$ & $(\bar{3},1,\alpha\!-\!\frac{2}{3})^-$ 
		&  $\bar{S}_1$& $R_2^{\dagger}$ & $[X_2,X_3]|_{\alpha=0}$\\
		B-2     & $u_R$  & $(1,2,\alpha)^-$ & $(1,2,\alpha)^-$ & $(\bar{3},2,\alpha\!-\!\frac{2}{3})^-$ & 
		$\bar{S}_1$ & $R_2^{\dagger}$ & $X_3|_{\alpha=\frac{1}{2}}$\\
		C-1     & $d_R$ & $(1,1\oplus3,\alpha)^-$ & $(1,1\oplus3,\alpha)^-$ & $(\bar{3},1\oplus3,\alpha\!+\!\frac{1}{3})^-$ & 
		$S_{1,3}$ & $\widetilde{R}_2^{\dagger}$ & $[X_2,X_3]|_{\alpha=0}$\\
		C-2     & $d_R$  & $(1,2,\alpha)^-$ & $(1,2,\alpha)^-$ & $(\bar{3},2,\alpha\!+\!\frac{1}{3})^-$ & 
		$S_{1,3}$ & $\widetilde{R}_2^{\dagger}$ & $X_3|_{\alpha=\frac{1}{2}}$\\
		D-1     & $Q_L$ & $(1,1\oplus3,\alpha)^-$ & $(1,1\oplus3,\alpha)^-$ & $(\bar{3},2,\alpha\!-\!\frac{1}{6})^-$ 
		& $\widetilde{V}_{2}$ & $U_{1,3}^{\dagger}$ & $[X_2,X_3]|_{\alpha=0}$\\
		D-2     & $Q_L$ & $(1,2,\alpha)^-$ & $(1,2,\alpha)^-$ & $(\bar{3},1\oplus3,\alpha\!-\!\frac{1}{6})^-$ 
		& $\widetilde{V}_{2}$ & $U_{1,3}^{\dagger}$ & $X_3|_{\alpha=\frac{1}{2}}$ \\
		E-1     & $\bar{u}_R$ & $(1,1\oplus3,\alpha)^-$ & $(1,1\oplus3,\alpha)^-$ & $(3,1\oplus3,\alpha\!+\!\frac{2}{3})^-$ 
		& $U_{1,3}$ & $\widetilde{V}_2^{\dagger}$ & $[X_2,X_3]|_{\alpha=0}$\\
		E-2     & $\bar{u}_R$  & $(1,2,\alpha)^-$ & $(1,2,\alpha)^-$ & $(3,2,\alpha\!+\!\frac{2}{3})^-$ 
		& $U_{1,3}$ & $\widetilde{V}_2^{\dagger}$ & $X_3|_{\alpha=\frac{1}{2}}$\\
		F-1     & $\bar{d}_R$ & $(1,1\oplus3,\alpha)^-$ & $(1,1\oplus3,\alpha)^-$ & $(3,1\oplus3,\alpha\!-\!\frac{1}{3})^-$ 
		& $\bar{U}_{1}$ & $V_2^{\dagger}$ & $[X_2,X_3]|_{\alpha=0}$\\
		F-2     & $\bar{d}_R$  & $(1,2,\alpha)^-$ & $(1,2,\alpha)^-$ & $(3,2,\alpha\!-\!\frac{1}{3})^-$ 
		& $\bar{U}_{1}$ & $V_2^{\dagger}$ & $X_3|_{\alpha=\frac{1}{2}}$\\
		\hline\hline		
	\end{tabular}
	\caption{Same as in Table~\ref{tab:T1-i-a} but for the diagram T3-ii depicted in Fig.~\ref{fig:4d2loop} and with the combination (a)-(A) shown in Fig.~\ref{fig:Z2}.}
	\label{tab:T3-ii-a}   
\end{table*}

\begin{table*}[htbp] 
	\renewcommand\arraystretch{1.5}  
	\centering 
	\tabcolsep=0.265cm
	\begin{tabular}{cccccccc}
		\hline\hline
		Model & $X^F_1$ & $X^F_2$ & $X_3$ & $X_4$  & $X_5$ & $X_6$  & DM \\ \hline
		A-1     & $(1,1\oplus3,\alpha)^-$  & $\bar{d}_R$ & $S_{1}$ & $(\bar{3},1\oplus3,\frac{1}{3}\!-\!\alpha)^-$ 
		& $(1,1\oplus3,-\alpha)^-$ & $(1,2,-\frac{1}{2}\!-\!\alpha)^-$ & $[X_1,X_5,X_6]|_{\alpha=0}$\\
		A-2     & $(1,2,\alpha)^-$  & $\bar{d}_R$ & $S_{1}$ & $(\bar{3},2,\frac{1}{3}\!-\!\alpha)^-$ 
		& $(1,2,-\alpha)^-$ & $(1,1\oplus3,-\frac{1}{2}\!-\!\alpha)^-$ & $[X_5,X_6]|_{\alpha=-\frac{1}{2}}$\\ 
		B-1     & $(1,1\oplus3,\alpha)^-$  & $Q_L$ & $\widetilde{R}_{2}$ & $(3,2,\frac{1}{6}\!-\!\alpha)^-$ 
		& $(1,1\oplus3,-\alpha)^-$ & $(1,2,-\frac{1}{2}\!-\!\alpha)^-$ & $[X_1,X_5,X_6]|_{\alpha=0}$\\
		B-2     & $(1,2,\alpha)^-$  & $Q_L$ & $\widetilde{R}_{2}$ & $(3,1\oplus3,\frac{1}{6}\!-\!\alpha)^-$ 
		& $(1,2,-\alpha)^-$ & $(1,1\oplus3,-\frac{1}{2}\!-\!\alpha)^-$ & $[X_5,X_6]|_{\alpha=-\frac{1}{2}}$\\ 
		C-1     & $(1,1\oplus3,\alpha)^-$  & $\bar{u}_R$ & $\bar{S}_{1}$ & $(\bar{3},1\oplus3,-\frac{2}{3}\!-\!\alpha)^-$ 
		& $(1,1\oplus3,-\alpha)^-$ & $(1,2,-\frac{1}{2}\!-\!\alpha)^-$ & $[X_1,X_5,X_6]|_{\alpha=0}$\\
		C-2     & $(1,2,\alpha)^-$  & $\bar{u}_R$ & $\bar{S}_{1}$ & $(\bar{3},2,-\frac{2}{3}\!-\!\alpha)^-$ 
		& $(1,2,-\alpha)^-$ & $(1,1\oplus3,-\frac{1}{2}\!-\!\alpha)^-$ & $[X_5,X_6]|_{\alpha=-\frac{1}{2}}$\\ 
		D-1     & $(1,1\oplus3,\alpha)^-$  & $u_R$ & $U_{1}$ & $(3,1\oplus3,\frac{2}{3}\!-\!\alpha)^-$ 
		& $(1,1\oplus3,-\alpha)^-$ & $(1,2,-\frac{1}{2}\!-\!\alpha)^-$ & $[X_1,X_5,X_6]|_{\alpha=0}$\\
		D-2     & $(1,2,\alpha)^-$  & $u_R$ & $U_{1}$ & $(3,2,\frac{2}{3}\!-\!\alpha)^-$ 
		& $(1,2,-\alpha)^-$ & $(1,1\oplus3,-\frac{1}{2}\!-\!\alpha)^-$ & $[X_5,X_6]|_{\alpha=-\frac{1}{2}}$\\ 
		E-1     & $(1,1\oplus3,\alpha)^-$  & $\bar{Q}_L$ & $\widetilde{V}_{2}$ & $(\bar{3},2,-\frac{1}{6}\!-\!\alpha)^-$ 
		& $(1,1\oplus3,-\alpha)^-$ & $(1,2,-\frac{1}{2}\!-\!\alpha)^-$ & $[X_1,X_5,X_6]|_{\alpha=0}$\\
		E-2     & $(1,2,\alpha)^-$  & $\bar{Q}_L$ & $\widetilde{V}_{2}$ & $(\bar{3},1\oplus3,-\frac{1}{6}\!-\!\alpha)^-$ 
		& $(1,2,-\alpha)^-$ & $(1,1\oplus3,-\frac{1}{2}\!-\!\alpha)^-$ & $[X_5,X_6]|_{\alpha=-\frac{1}{2}}$\\ 
		F-1     & $(1,1\oplus3,\alpha)^-$  & $d_R$ & $\bar{U}_{1}$ & $(3,1\oplus3,-\frac{1}{3}\!-\!\alpha)^-$ 
		& $(1,1\oplus3,-\alpha)^-$ & $(1,2,-\frac{1}{2}\!-\!\alpha)^-$ & $[X_1,X_5,X_6]|_{\alpha=0}$\\
		F-2     & $(1,2,\alpha)^-$  & $d_R$ & $\bar{U}_{1}$ & $(3,2,-\frac{1}{3}\!-\!\alpha)^-$ 
		& $(1,2,-\alpha)^-$ & $(1,1\oplus3,-\frac{1}{2}\!-\!\alpha)^-$ & $[X_5,X_6]|_{\alpha=-\frac{1}{2}}$\\ 
		\hline\hline		
	\end{tabular}
	\caption{Same as in Table~\ref{tab:T1-i-a} but for the diagram T3-ii depicted in Fig.~\ref{fig:4d2loop} and with the combination (b)-(B) shown in Fig.~\ref{fig:Z2}.}
	\label{tab:T3-ii-b}   
\end{table*}

\begin{table*}[htbp] 
	\renewcommand\arraystretch{1.5}  
	\centering 
	\tabcolsep=0.335cm
	\begin{tabular}{cccccccc}
		\hline\hline
		Model & $X^F_1$ & $X^F_2$ & $X_3$ & $X_4$  & $X_5$ & $X_6$ & DM \\ \hline
		A-1     & $\bar{Q}_L$ & $(1,1\oplus3,\alpha)^-$ & $(1,1\oplus3,\alpha)^-$ &$(3,1\oplus3,\alpha\!-\!\frac{1}{3})^-$ & $(3,2,\alpha\!+\!\frac{1}{6})^-$ & $S_{1,3}^{\dagger}$ & $[X_2,X_3]|_{\alpha=0}$\\
		A-2     & $\bar{Q}_L$ & $(1,2,\alpha)^-$ & $(1,2,\alpha)^-$ &$(3,2,\alpha\!-\!\frac{1}{3})^-$  & $(3,1\oplus3,\alpha\!+\!\frac{1}{6})^-$ & $S_{1,3}^{\dagger}$ & $X_3|_{\alpha=\frac{1}{2}}$\\
		B-1     & $u_R$ & $(1,1\oplus3,\alpha)^-$ & $(1,1\oplus3,\alpha)^-$ &$(\bar{3},2,\alpha\!-\!\frac{7}{6})^-$ & $(\bar{3},1,\alpha\!-\!\frac{2}{3})^-$ & $R_2^{\dagger}$ & $[X_2,X_3]|_{\alpha=0}$\\
		B-2     & $u_R$  & $(1,2,\alpha)^-$ & $(1,2,\alpha)^-$& $(\bar{3},1\oplus3,\alpha\!-\!\frac{7}{6})^-$ & $(\bar{3},2,\alpha\!-\!\frac{2}{3})^-$ & $R_2^{\dagger}$ & $X_3|_{\alpha=\frac{1}{2}}$\\
		C-1     & $d_R$ & $(1,1\oplus3,\alpha)^-$ & $(1,1\oplus3,\alpha)^-$ &$(\bar{3},2,\alpha\!-\!\frac{1}{6})^-$ & $(\bar{3},1\oplus3,\alpha\!+\!\frac{1}{3})^-$ & $\widetilde{R}_2^{\dagger}$ & $[X_2,X_3]|_{\alpha=0}$\\
		C-2     & $d_R$  & $(1,2,\alpha)^-$ & $(1,2,\alpha)^-$& $(\bar{3},1\oplus3,\alpha\!-\!\frac{1}{6})^-$  & $(\bar{3},2,\alpha\!+\!\frac{1}{3})^-$ & $\widetilde{R}_2^{\dagger}$ & $X_3|_{\alpha=\frac{1}{2}}$\\
		D-1     & $Q_L$ & $(1,1\oplus3,\alpha)^-$ & $(1,1\oplus3,\alpha)^-$ &$(\bar{3},1\oplus3,\alpha\!-\!\frac{2}{3})^-$  & $(\bar{3},2,\alpha\!-\!\frac{1}{6})^-$ & $U_{1,3}^{\dagger}$ & $[X_2,X_3]|_{\alpha=0}$\\
		D-2     & $Q_L$ & $(1,2,\alpha)^-$ & $(1,2,\alpha)^-$ & $(\bar{3},2,\alpha\!-\!\frac{2}{3})^-$  & $(\bar{3},1\oplus3,\alpha\!-\!\frac{1}{6})^-$ & $U_{1,3}^{\dagger}$ &  $X_3|_{\alpha=\frac{1}{2}}$\\
		E-1     & $\bar{u}_R$ & $(1,1\oplus3,\alpha)^-$ & $(1,1\oplus3,\alpha)^-$&$(3,1\oplus3,\alpha\!+\!\frac{1}{6})^-$  & $(3,1\oplus3,\alpha\!+\!\frac{2}{3})^-$ & $\widetilde{V}_2^{\dagger}$ & $[X_2,X_3]|_{\alpha=0}$\\
		E-2     & $\bar{u}_R$  & $(1,2,\alpha)^-$ & $(1,2,\alpha)^-$ & $(3,2,\alpha\!+\!\frac{1}{6})^-$ &$(3,2,\alpha\!+\!\frac{2}{3})^-$ & $\widetilde{V}_2^{\dagger}$ & $X_3|_{\alpha=\frac{1}{2}}$\\
		F-1     & $\bar{d}_R$ & $(1,1\oplus3,\alpha)^-$ & $(1,1\oplus3,\alpha)^-$&$(3,1\oplus3,\alpha\!-\!\frac{5}{6})^-$  & $(3,1\oplus3,\alpha\!-\!\frac{1}{3})^-$ & $V_2^{\dagger}$ & $[X_2,X_3]|_{\alpha=0}$\\
		F-2     & $\bar{d}_R$  & $(1,2,\alpha)^-$ & $(1,2,\alpha)^-$ & $(3,2,\alpha\!-\!\frac{5}{6})^-$ & $(3,2,\alpha\!-\!\frac{1}{3})^-$ & $V_2^{\dagger}$ &$X_3|_{\alpha=\frac{1}{2}}$\\
		\hline\hline		
	\end{tabular}
	\caption{Same as in Table~\ref{tab:T1-i-a} but for the diagram T2-i depicted in Fig.~\ref{fig:4d2loop} and with the combination (a)-(A) shown in Fig.~\ref{fig:Z2}.}
	\label{tab:T2-i-a}   
\end{table*}

\begin{table*}[htbp] 
	\renewcommand\arraystretch{1.5}  
	\centering 
	\tabcolsep=0.265cm
	\begin{tabular}{cccccccc}
		\hline\hline
		Model & $X^F_1$ & $X^F_2$ & $X_3$ & $X_4$  & $X_5$ & $X_6$ & DM \\ \hline
		A-1     & $(1,1\oplus3,\alpha)^-$  & $\bar{d}_R$ & $S_{1}$ & $(\bar{3},2,-\frac{1}{6}\!-\!\alpha)^-$  & $(\bar{3},1\oplus3,\frac{1}{3}\!-\!\alpha)^-$ & $(1,2,-\frac{1}{2}\!-\!\alpha)^-$ & $[X_1,X_6]|_{\alpha=0}$\\
		A-2     & $(1,2,\alpha)^-$  & $\bar{d}_R$ & $S_{1}$ &  $(\bar{3},1\oplus3,-\frac{1}{6}\!-\!\alpha)^-$& 
		$(\bar{3},2,\frac{1}{3}\!-\!\alpha)^-$ & $(1,1\oplus3,-\frac{1}{2}\!-\!\alpha)^-$ &$X_6|_{\alpha=-\frac{1}{2}}$\\ 
		B-1     & $(1,1\oplus3,\alpha)^-$  & $Q_L$ & $\widetilde{R}_{2}$ & $(3,1\oplus3,-\frac{1}{3}\!-\!\alpha)^-$ 
		& $(3,2,\frac{1}{6}\!-\!\alpha)^-$& $(1,2,-\frac{1}{2}\!-\!\alpha)^-$ & $[X_1,X_6]|_{\alpha=0}$\\
		B-2     & $(1,2,\alpha)^-$  & $Q_L$ & $\widetilde{R}_{2}$& $(3,2,-\frac{1}{3}\!-\!\alpha)^-$ 
		& $(3,1\oplus3,\frac{1}{6}\!-\!\alpha)^-$ & $(1,1\oplus3,-\frac{1}{2}\!-\!\alpha)^-$ & $X_6|_{\alpha=-\frac{1}{2}}$\\ 
		C-1     & $(1,1\oplus3,\alpha)^-$  & $\bar{u}_R$ & $\bar{S}_{1}$ & $(\bar{3},2,-\frac{7}{6}\!-\!\alpha)^-$ &
		$(\bar{3},1\oplus3,-\frac{2}{3}\!-\!\alpha)^-$ & $(1,2,-\frac{1}{2}\!-\!\alpha)^-$ & $[X_1,X_6]|_{\alpha=0}$\\
		C-2     & $(1,2,\alpha)^-$  & $\bar{u}_R$ & $\bar{S}_{1}$ & $(\bar{3},1\oplus3,-\frac{7}{6}\!-\!\alpha)^-$ &
		$(\bar{3},2,-\frac{2}{3}\!-\!\alpha)^-$ & $(1,1\oplus3,-\frac{1}{2}\!-\!\alpha)^-$ & $X_6|_{\alpha=-\frac{1}{2}}$\\
		D-1     & $(1,1\oplus3,\alpha)^-$  & $u_R$ & $U_{1}$ & $(3,2,\frac{1}{6}\!-\!\alpha)^-$ & $(3,1\oplus3,\frac{2}{3}\!-\!\alpha)^-$ & $(1,2,-\frac{1}{2}\!-\!\alpha)^-$ & $[X_1,X_6]|_{\alpha=0}$\\
		D-2     & $(1,2,\alpha)^-$  & $u_R$ & $U_{1}$ & $(3,1\oplus3,\frac{1}{6}\!-\!\alpha)^-$ &$(3,2,\frac{2}{3}\!-\!\alpha)^-$ & $(1,1\oplus3,-\frac{1}{2}\!-\!\alpha)^-$ & $X_6|_{\alpha=-\frac{1}{2}}$\\
		E-1     & $(1,1\oplus3,\alpha)^-$  & $\bar{Q}_L$ & $\widetilde{V}_{2}$ &$(\bar{3},1\oplus3,-\frac{2}{3}\!-\!\alpha)^-$  & $(\bar{3},2,-\frac{1}{6}\!-\!\alpha)^-$ & $(1,2,-\frac{1}{2}\!-\!\alpha)^-$ & $[X_1,X_6]|_{\alpha=0}$\\
		E-2     & $(1,2,\alpha)^-$  & $\bar{Q}_L$ & $\widetilde{V}_{2}$ & $(\bar{3},2,-\frac{2}{3}\!-\!\alpha)^-$& $(\bar{3},1\oplus3,-\frac{1}{6}\!-\!\alpha)^-$ & $(1,1\oplus3,-\frac{1}{2}\!-\!\alpha)^-$ & $X_6|_{\alpha=-\frac{1}{2}}$\\
		F-1     & $(1,1\oplus3,\alpha)^-$  & $d_R$ & $\bar{U}_{1}$ & $(3,2,-\frac{5}{6}\!-\!\alpha)^-$ & $(3,1\oplus3,-\frac{1}{3}\!-\!\alpha)^-$ & $(1,2,-\frac{1}{2}\!-\!\alpha)^-$ & $[X_1,X_6]|_{\alpha=0}$\\
		F-2     & $(1,2,\alpha)^-$  & $d_R$ & $\bar{U}_{1}$ & $(3,1\oplus3,-\frac{5}{6}\!-\!\alpha)^-$  & $(3,2,-\frac{1}{3}\!-\!\alpha)^-$ & $(1,1\oplus3,-\frac{1}{2}\!-\!\alpha)^-$ & $X_6|_{\alpha=-\frac{1}{2}}$\\
		\hline\hline		
	\end{tabular}
	\caption{Same as in Table~\ref{tab:T1-i-a} but for the diagram T2-i depicted in Fig.~\ref{fig:4d2loop} and with the combination (b)-(B) shown in Fig.~\ref{fig:Z2}.}
	\label{tab:T2-i-b}   
\end{table*}

\begin{table*}[htbp] 
	\renewcommand\arraystretch{1.5}  
	\centering 
	\tabcolsep=0.415cm
	\begin{tabular}{cccccccc}
		\hline\hline
		Model & $X^F_1$ & $X^F_2$ & $X^F_3$ & $X_4$  & $X_5$ & $X_6$ & DM \\ \hline
		A-1     & $\bar{Q}_L$ &$\bar{d}_R$ &$(1,1\oplus3,\alpha)^-$ & $(1,1\oplus3,\alpha)^-$ & $(3,1\oplus3,\alpha\!-\!\frac{1}{3})^-$ & $S_{1,3}^{\dagger}$ & $[X_3,X_4]|_{\alpha=0}$\\
		A-2     & $\bar{Q}_L$ &$\bar{d}_R$ & $(1,2,\alpha)^-$ & $(1,2,\alpha)^-$ & $(3,2,\alpha\!-\!\frac{1}{3})^-$ 
		& $S_{1,3}^{\dagger}$ & $X_4|_{\alpha=\frac{1}{2}}$\\
		B-1     & $u_R$ & $(3,2,\frac{7}{6})$& $(1,1\oplus3,\alpha)^-$ & $(1,1\oplus3,\alpha)^-$ & $(\bar{3},2,\alpha\!-\!\frac{7}{6})^-$ & $R_2^{\dagger}$ & $[X_3,X_4]|_{\alpha=0}$\\
		B-2     & $u_R$ & $(3,2,\frac{7}{6})$ & $(1,2,\alpha)^-$ & $(1,2,\alpha)^-$& $(\bar{3},1\oplus3,\alpha\!-\!\frac{7}{6})^-$ & $R_2^{\dagger}$ &  $X_4|_{\alpha=\frac{1}{2}}$\\
		C-1     & $d_R$ & $Q_L$ & $(1,1\oplus3,\alpha)^-$ & $(1,1\oplus3,\alpha)^-$ & 
		$(\bar{3},2,\alpha\!-\!\frac{1}{6})^-$ & $\widetilde{R}_2^{\dagger}$ & $[X_3,X_4]|_{\alpha=0}$\\
		C-2     & $d_R$  & $Q_L$ & $(1,2,\alpha)^-$ & $(1,2,\alpha)^-$& $(\bar{3},1\oplus3,\alpha\!-\!\frac{1}{6})^-$  & $\widetilde{R}_2^{\dagger}$ &  $X_4|_{\alpha=\frac{1}{2}}$\\
		D-1     & $Q_L$ & $u_R$& $(1,1\oplus3,\alpha)^-$ & $(1,1\oplus3,\alpha)^-$ &$(\bar{3},1\oplus3,\alpha\!-\!\frac{2}{3})^-$  & $U_{1,3}^{\dagger}$ &$[X_3,X_4]|_{\alpha=0}$\\
		D-2     & $Q_L$ & $u_R$ & $(1,2,\alpha)^-$ & $(1,2,\alpha)^-$ & $(\bar{3},2,\alpha\!-\!\frac{2}{3})^-$  & $U_{1,3}^{\dagger}$ &   $X_4|_{\alpha=\frac{1}{2}}$\\
		E-1     & $\bar{u}_R$ & $\bar{Q}_L$  & $(1,1\oplus3,\alpha)^-$ & $(1,1\oplus3,\alpha)^-$&$(3,2,\alpha\!+\!\frac{1}{6})^-$  &  $\widetilde{V}_2^{\dagger}$ & $[X_3,X_4]|_{\alpha=0}$\\
		E-2     & $\bar{u}_R$   & $\bar{Q}_L$  & $(1,2,\alpha)^-$ & $(1,2,\alpha)^-$ & $(3,1\oplus3,\alpha\!+\!\frac{1}{6})^-$ & $\widetilde{V}_2^{\dagger}$ & $X_4|_{\alpha=\frac{1}{2}}$\\
		F-1     & $\bar{d}_R$ &$(\bar{3},2,\frac{5}{6})$& $(1,1\oplus3,\alpha)^-$ & $(1,1\oplus3,\alpha)^-$&$(3,2,\alpha\!-\!\frac{5}{6})^-$  & $V_2^{\dagger}$ & $[X_3,X_4]|_{\alpha=0}$\\
		F-2     & $\bar{d}_R$  &$(\bar{3},2,\frac{5}{6})$ & $(1,2,\alpha)^-$ & $(1,2,\alpha)^-$ & $(3,1\oplus3,\alpha\!-\!\frac{5}{6})^-$ & $V_2^{\dagger}$ & $X_4|_{\alpha=\frac{1}{2}}$\\
		\hline\hline		
	\end{tabular}
	\caption{Same as in Table~\ref{tab:T1-i-a} but for the diagram T3-iii depicted in Fig.~\ref{fig:4d2loop} and with the combination (a)-(A) shown in Fig.~\ref{fig:Z2}.}
	\label{tab:T3-iii-a}   
\end{table*}

\begin{table*}[htbp] 
	\renewcommand\arraystretch{1.5}  
	\centering 
	\tabcolsep=0.255cm
	\begin{tabular}{cccccccc}
		\hline\hline
		Model & $X^F_1$ & $X^F_2$ & $X^F_3$ & $X_4$  & $X_5$ & $X_6$ & DM \\ \hline
		A-1     & $(1,1\oplus3,\alpha)^-$  & $(1,2,\alpha+\frac{1}{2})^-$  & $\bar{d}_R$ & $S_{1}$ & $(\bar{3},2,-\frac{1}{6}\!-\!\alpha)^-$  & $(1,2,-\frac{1}{2}\!-\!\alpha)^-$ & $[X_1,X_6]|_{\alpha=0}$\\
		A-2     & $(1,2,\alpha)^-$ & $(1,1\oplus3,\alpha+\frac{1}{2})^-$ & $\bar{d}_R$ & $S_{1}$ &  $(\bar{3},1\oplus3,-\frac{1}{6}\!-\!\alpha)^-$&$(1,1\oplus3,-\frac{1}{2}\!-\!\alpha)^-$ &$[X_2,X_6]|_{\alpha=-\frac{1}{2}}$\\ 
		B-1     & $(1,1\oplus3,\alpha)^-$ & $(1,2,\alpha+\frac{1}{2})^-$ & $Q_L$ & $\widetilde{R}_{2}$ & $(3,1\oplus3,-\frac{1}{3}\!-\!\alpha)^-$ & $(1,2,-\frac{1}{2}\!-\!\alpha)^-$ & $[X_1,X_6]|_{\alpha=0}$\\
		B-2     & $(1,2,\alpha)^-$ & $(1,1\oplus3,\alpha+\frac{1}{2})^-$  & $Q_L$ & $\widetilde{R}_{2}$& $(3,2,-\frac{1}{3}\!-\!\alpha)^-$ 
		& $(1,1\oplus3,-\frac{1}{2}\!-\!\alpha)^-$ & $[X_2,X_6]|_{\alpha=-\frac{1}{2}}$\\ 
		C-1     & $(1,1\oplus3,\alpha)^-$ & $(1,2,\alpha+\frac{1}{2})^-$ & $\bar{u}_R$ & $\bar{S}_{1}$ & $(\bar{3},2,-\frac{7}{6}\!-\!\alpha)^-$ &$(1,2,-\frac{1}{2}\!-\!\alpha)^-$ & $[X_1,X_6]|_{\alpha=0}$\\
		C-2     & $(1,2,\alpha)^-$  & $(1,1\oplus3,\alpha+\frac{1}{2})^-$ & $\bar{u}_R$ & $\bar{S}_{1}$ & $(\bar{3},1\oplus3,-\frac{7}{6}\!-\!\alpha)^-$ &
		$(1,1\oplus3,-\frac{1}{2}\!-\!\alpha)^-$ & $[X_2,X_6]|_{\alpha=-\frac{1}{2}}$\\ 
		D-1     & $(1,1\oplus3,\alpha)^-$  & $(1,2,\alpha+\frac{1}{2})^-$   & $u_R$ & $U_{1}$ & $(3,2,\frac{1}{6}\!-\!\alpha)^-$ & $(1,2,-\frac{1}{2}\!-\!\alpha)^-$ & $[X_1,X_6]|_{\alpha=0}$\\
		D-2     & $(1,2,\alpha)^-$  & $(1,1\oplus3,\alpha+\frac{1}{2})^-$ & $u_R$ & $U_{1}$ & $(3,1\oplus3,\frac{1}{6}\!-\!\alpha)^-$  & $(1,1\oplus3,-\frac{1}{2}\!-\!\alpha)^-$ &$[X_2,X_6]|_{\alpha=-\frac{1}{2}}$\\ 
		E-1     & $(1,1\oplus3,\alpha)^-$ & $(1,2,\alpha+\frac{1}{2})^-$   & $\bar{Q}_L$ & $\widetilde{V}_{2}$ &$(\bar{3},1\oplus3,-\frac{2}{3}\!-\!\alpha)^-$  & $(1,2,-\frac{1}{2}\!-\!\alpha)^-$ & $[X_1,X_6]|_{\alpha=0}$\\
		E-2     & $(1,2,\alpha)^-$ & $(1,1\oplus3,\alpha+\frac{1}{2})^-$ & $\bar{Q}_L$ & $\widetilde{V}_{2}$ & $(\bar{3},2,-\frac{2}{3}\!-\!\alpha)^-$ & $(1,1\oplus3,-\frac{1}{2}\!-\!\alpha)^-$ & $[X_2,X_6]|_{\alpha=-\frac{1}{2}}$\\ 
		F-1     & $(1,1\oplus3,\alpha)^-$ & $(1,2,\alpha+\frac{1}{2})^-$  &  $d_R$ & $\bar{U}_{1}$ & $(3,2,-\frac{5}{6}\!-\!\alpha)^-$ &  $(1,2,-\frac{1}{2}\!-\!\alpha)^-$ & $[X_1,X_6]|_{\alpha=0}$\\
		F-2     & $(1,2,\alpha)^-$ & $(1,1\oplus3,\alpha+\frac{1}{2})^-$ & $d_R$ & $\bar{U}_{1}$ & $(3,1\oplus3,-\frac{5}{6}\!-\!\alpha)^-$  & $(1,1\oplus3,-\frac{1}{2}\!-\!\alpha)^-$ & $[X_2,X_6]|_{\alpha=-\frac{1}{2}}$\\ 
		\hline\hline		
	\end{tabular}
	\caption{Same as in Table~\ref{tab:T1-i-a} but for the diagram T3-iii depicted in Fig.~\ref{fig:4d2loop} and with the combination (b)-(B) shown in Fig.~\ref{fig:Z2}.}
	\label{tab:T3-iii-b}   
\end{table*}

\begin{table*}[htbp] 
	\renewcommand\arraystretch{1.5}  
	\centering 
	\tabcolsep=0.44cm
	\begin{tabular}{cccccccc}
		\hline\hline
		Model & $X_1$ & $X^F_2$ & $X^F_3$ & $X^F_4$  & $X_5$ & $X^F_6$ & DM \\ \hline
		A-1  & $S_{1}^{\dagger}$   &$d_R$ & $(1,2,\alpha)^-$ & $(1,1\oplus3,\alpha+\frac{1}{2})^-$ & $(\bar{3},2,\alpha\!+\!\frac{1}{3})^-$ &  $\bar{Q}_L$ & $X_4|_{\alpha=-\frac{1}{2}}$\\
		A-2    & $S_{1}^{\dagger}$   &$d_R$ & $(1,1\oplus3,\alpha)^-$ & $(1,2,\alpha+\frac{1}{2})^-$ & $(\bar{3},1\oplus3,\alpha\!+\!\frac{1}{3})^-$ &  $\bar{Q}_L$ & $X_3|_{\alpha=0}$\\
		B-1     & $\widetilde{R}_2^{\dagger}$ &$\bar{Q}_L$   & $(1,2,\alpha)^-$ & $(1,1\oplus3,\alpha+\frac{1}{2})^-$ & $(3,1\oplus3,\alpha+\frac{1}{6})^-$ &   $d_R$ & $X_4|_{\alpha=-\frac{1}{2}}$\\
		B-2     & $\widetilde{R}_2^{\dagger}$ &$\bar{Q}_L$   & $(1,1\oplus3,\alpha)^-$ & $(1,2,\alpha+\frac{1}{2})^-$ & $(3,2,\alpha+\frac{1}{6})^-$ &   $d_R$ &  $X_3|_{\alpha=0}$\\	
		C-1  & $U_{1}^{\dagger}$   &$\bar{u}_R$ & $(1,2,\alpha)^-$ & $(1,1\oplus3,\alpha+\frac{1}{2})^-$ & $(3,2,\alpha\!+\!\frac{2}{3})^-$ &  $Q_L$ & $X_4|_{\alpha=-\frac{1}{2}}$\\
		C-2    & $U_{1}^{\dagger}$   &$\bar{u}_R$ & $(1,1\oplus3,\alpha)^-$ & $(1,2,\alpha+\frac{1}{2})^-$ & $(3,1\oplus3,\alpha\!+\!\frac{2}{3})^-$ &  $Q_L$ & $X_3|_{\alpha=0}$\\		
		D-1    & $\widetilde{V}_2^{\dagger}$ & $Q_L$   & $(1,2,\alpha)^-$ & $(1,1\oplus3,\alpha+\frac{1}{2})^-$ & $(\bar{3},1\oplus3,\alpha\!-\!\frac{1}{6})^-$ & $\bar{u}_R$ & $X_4|_{\alpha=-\frac{1}{2}}$\\
		D-2  & $\widetilde{V}_2^{\dagger}$   &$Q_L$& $(1,1\oplus3,\alpha)^-$ & $(1,2,\alpha+\frac{1}{2})^-$ & $(\bar{3},2,\alpha\!-\!\frac{1}{6})^-$ &  $\bar{u}_R$   &$X_3|_{\alpha=0}$\\		
		\hline\hline		
	\end{tabular}
	\caption{Same as in Table~\ref{tab:T1-i-a} but for the diagram T3-ix depicted in Fig.~\ref{fig:4d2loop} and with the combination (a)-(A) shown in Fig.~\ref{fig:Z2}.}
	\label{tab:T3-ix-a}   
\end{table*}

\begin{table*}[htbp] 
	\renewcommand\arraystretch{1.5}  
	\centering 
	\tabcolsep=0.275cm
	\begin{tabular}{cccccccc}
		\hline\hline
		Model & $X^F_1$ & $X^F_2$ & $X^F_3$ & $X_4$  & $X_5$ & $X_6$ & DM \\ \hline
		A-1     & $\bar{Q}_L$ &$(1,1\oplus3,\alpha)^-$ & $(1,2,\alpha+\frac{1}{2})^-$ & $(1,2,\alpha+\frac{1}{2})^-$ & $(3,2,\alpha\!+\!\frac{1}{6})^-$ & $S_{1,3}^{\dagger}$ & $[X_2,X_4]|_{\alpha=0}$\\
		A-2     & $\bar{Q}_L$ & $(1,2,\alpha)^-$ & $(1,1\oplus3,\alpha+\frac{1}{2})^-$ & $(1,1\oplus3,\alpha+\frac{1}{2})^-$
		& $(3,1\oplus3,\alpha\!+\!\frac{1}{6})^-$ 
		& $S_{1,3}^{\dagger}$ & $[X_3,X_4]|_{\alpha=-\frac{1}{2}}$\\
		B-1     & $u_R$ & $(1,1\oplus3,\alpha)^-$ & $(1,2,\alpha+\frac{1}{2})^-$ & $(1,2,\alpha+\frac{1}{2})^-$ & $(\bar{3},2,\alpha\!-\!\frac{2}{3})^-$ & $R_2^{\dagger}$ & $[X_2,X_4]|_{\alpha=0}$\\
		B-2     & $u_R$ & $(1,2,\alpha)^-$ & $(1,1\oplus3,\alpha+\frac{1}{2})^-$ & $(1,1\oplus3,\alpha+\frac{1}{2})^-$& $(\bar{3},1\oplus3,\alpha\!-\!\frac{2}{3})^-$ & $R_2^{\dagger}$ &  $[X_3,X_4]|_{\alpha=-\frac{1}{2}}$\\
		C-1     & $d_R$ & $(1,1\oplus3,\alpha)^-$ & $(1,2,\alpha+\frac{1}{2})^-$ & $(1,2,\alpha+\frac{1}{2})^-$ &
		$(\bar{3},2,\alpha\!+\!\frac{1}{3})^-$ & $\widetilde{R}_2^{\dagger}$ & $[X_2,X_4]|_{\alpha=0}$\\
		C-2     & $d_R$  & $(1,2,\alpha)^-$ & $(1,1\oplus3,\alpha+\frac{1}{2})^-$ & $(1,1\oplus3,\alpha+\frac{1}{2})^-$ & $(\bar{3},1\oplus3,\alpha\!+\!\frac{1}{3})^-$  & $\widetilde{R}_2^{\dagger}$ &  $[X_3,X_4]|_{\alpha=-\frac{1}{2}}$\\
		D-1     & $Q_L$ & $(1,1\oplus3,\alpha)^-$ & $(1,2,\alpha+\frac{1}{2})^-$ & $(1,2,\alpha+\frac{1}{2})^-$ &$(\bar{3},1\oplus3,\alpha\!-\!\frac{1}{6})^-$  & $U_{1,3}^{\dagger}$ &$[X_2,X_4]|_{\alpha=0}$\\
		D-2     & $Q_L$ & $(1,2,\alpha)^-$ & $(1,1\oplus3,\alpha+\frac{1}{2})^-$ & $(1,1\oplus3,\alpha+\frac{1}{2})^-$ & $(\bar{3},2,\alpha\!-\!\frac{1}{6})^-$  & $U_{1,3}^{\dagger}$ &   $[X_3,X_4]|_{\alpha=-\frac{1}{2}}$\\
		E-1     & $\bar{u}_R$ & $(1,1\oplus3,\alpha)^-$ & $(1,2,\alpha+\frac{1}{2})^-$ & $(1,2,\alpha+\frac{1}{2})^-$ & $(3,2,\alpha\!+\!\frac{2}{3})^-$  &  $\widetilde{V}_2^{\dagger}$ &$[X_2,X_4]|_{\alpha=0}$\\
		E-2     & $\bar{u}_R$   & $(1,2,\alpha)^-$ & $(1,1\oplus3,\alpha+\frac{1}{2})^-$ & $(1,1\oplus3,\alpha+\frac{1}{2})^-$ & $(3,1\oplus3,\alpha\!+\!\frac{2}{3})^-$ & $\widetilde{V}_2^{\dagger}$ & $[X_3,X_4]|_{\alpha=-\frac{1}{2}}$\\
		F-1     & $\bar{d}_R$ & $(1,1\oplus3,\alpha)^-$ & $(1,2,\alpha+\frac{1}{2})^-$ & $(1,2,\alpha+\frac{1}{2})^-$ &$(3,2,\alpha\!-\!\frac{1}{3})^-$  & $V_2^{\dagger}$ & $[X_2,X_4]|_{\alpha=0}$\\
		F-2     & $\bar{d}_R$  & $(1,2,\alpha)^-$ & $(1,1\oplus3,\alpha+\frac{1}{2})^-$ & $(1,1\oplus3,\alpha+\frac{1}{2})^-$ & $(3,1\oplus3,\alpha\!-\!\frac{1}{3})^-$ & $V_2^{\dagger}$ & $[X_3,X_4]|_{\alpha=-\frac{1}{2}}$\\
		\hline\hline		
	\end{tabular}
	\caption{Same as in Table~\ref{tab:T1-i-a} but for the diagram T3-iv depicted in Fig.~\ref{fig:4d2loop} and with the combination (a)-(A) shown in Fig.~\ref{fig:Z2}.}
	\label{tab:T3-iv-a}   
\end{table*}

\begin{table*}[htbp] 
	\renewcommand\arraystretch{1.5}  
	\centering 
	\tabcolsep=0.345cm
	\begin{tabular}{cccccccc}
		\hline\hline
		Model & $X^F_1$ & $X^F_2$ & $X^F_3$ & $X_4$  & $X_5$ & $X_6$ & DM \\ \hline
		A-1     & $(1,1\oplus3,\alpha)^-$  & $\bar{Q}_L$  & $\bar{d}_R$ & $S_{1}$ & $(\bar{3},2,-\frac{1}{6}\!-\!\alpha)^-$  & $(1,2,-\frac{1}{2}\!-\!\alpha)^-$ & $[X_1,X_6]|_{\alpha=0}$\\
		A-2     & $(1,2,\alpha)^-$ & $\bar{Q}_L$  & $\bar{d}_R$ & $S_{1}$ &  $(\bar{3},1\oplus3,-\frac{1}{6}\!-\!\alpha)^-$&$(1,1\oplus3,-\frac{1}{2}\!-\!\alpha)^-$ &$X_6|_{\alpha=-\frac{1}{2}}$\\ 
		B-1     & $(1,1\oplus3,\alpha)^-$ & $d_R$ & $Q_L$ & $\widetilde{R}_{2}$ & $(3,1\oplus3,-\frac{1}{3}\!-\!\alpha)^-$ & $(1,2,-\frac{1}{2}\!-\!\alpha)^-$ & $[X_1,X_6]|_{\alpha=0}$\\
		B-2     & $(1,2,\alpha)^-$ & $d_R$  & $Q_L$ & $\widetilde{R}_{2}$& $(3,2,-\frac{1}{3}\!-\!\alpha)^-$ 
		& $(1,1\oplus3,-\frac{1}{2}\!-\!\alpha)^-$ & $X_6|_{\alpha=-\frac{1}{2}}$\\ 
		C-1     & $(1,1\oplus3,\alpha)^-$ & $(\bar{3},2,-\frac{7}{6})$ & $\bar{u}_R$ & $\bar{S}_{1}$ & $(\bar{3},2,-\frac{7}{6}\!-\!\alpha)^-$ &$(1,2,-\frac{1}{2}\!-\!\alpha)^-$ & $[X_1,X_6]|_{\alpha=0}$\\
		C-2     & $(1,2,\alpha)^-$  &  $(\bar{3},2,-\frac{7}{6})$ & $\bar{u}_R$ & $\bar{S}_{1}$ & $(\bar{3},1\oplus3,-\frac{7}{6}\!-\!\alpha)^-$ &
		$(1,1\oplus3,-\frac{1}{2}\!-\!\alpha)^-$ & $X_6|_{\alpha=-\frac{1}{2}}$\\
		D-1     & $(1,1\oplus3,\alpha)^-$  & $Q_L$   & $u_R$ & $U_{1}$ & $(3,2,\frac{1}{6}\!-\!\alpha)^-$ & $(1,2,-\frac{1}{2}\!-\!\alpha)^-$ & $[X_1,X_6]|_{\alpha=0}$\\
		D-2     & $(1,2,\alpha)^-$  &  $Q_L$ & $u_R$ & $U_{1}$ & $(3,1\oplus3,\frac{1}{6}\!-\!\alpha)^-$  & $(1,1\oplus3,-\frac{1}{2}\!-\!\alpha)^-$ & $X_6|_{\alpha=-\frac{1}{2}}$\\
		E-1     & $(1,1\oplus3,\alpha)^-$ & $\bar{u}_R$   & $\bar{Q}_L$ & $\widetilde{V}_{2}$ &$(\bar{3},1\oplus3,-\frac{2}{3}\!-\!\alpha)^-$  & $(1,2,-\frac{1}{2}\!-\!\alpha)^-$ & $[X_1,X_6]|_{\alpha=0}$\\
		E-2     & $(1,2,\alpha)^-$ & $\bar{u}_R$ & $\bar{Q}_L$ & $\widetilde{V}_{2}$ & $(\bar{3},2,-\frac{2}{3}\!-\!\alpha)^-$ & $(1,1\oplus3,-\frac{1}{2}\!-\!\alpha)^-$ & $X_6|_{\alpha=-\frac{1}{2}}$\\
		F-1     & $(1,1\oplus3,\alpha)^-$ & $(3,2,-\frac{5}{6})$  &  $d_R$ & $\bar{U}_{1}$ & $(3,2,-\frac{5}{6}\!-\!\alpha)^-$ &  $(1,2,-\frac{1}{2}\!-\!\alpha)^-$ & $[X_1,X_6]|_{\alpha=0}$\\
		F-2     & $(1,2,\alpha)^-$ &  $(3,2,-\frac{5}{6})$  & $d_R$ & $\bar{U}_{1}$ & $(3,1\oplus3,-\frac{5}{6}\!-\!\alpha)^-$  & $(1,1\oplus3,-\frac{1}{2}\!-\!\alpha)^-$ & $X_6|_{\alpha=-\frac{1}{2}}$\\
		\hline\hline		
	\end{tabular}
	\caption{Same as in Table~\ref{tab:T1-i-a} but for the diagram T3-iv depicted in Fig.~\ref{fig:4d2loop} and with the combination (b)-(B) shown in Fig.~\ref{fig:Z2}.}
	\label{tab:T3-iv-b}   
\end{table*}

\begin{table*}[htbp] 
	\renewcommand\arraystretch{1.5}  
	\centering 
	\tabcolsep=0.395cm
	\begin{tabular}{cccccccc}
		\hline\hline
		Model & $X_1$ & $X^F_2$ & $X_3$ & $X_4$  & $X^F_5$ & $X^F_6$ & DM \\ \hline
		A-1  & $S_{1}^{\dagger}$   &$d_R$ & $(1,2,\alpha)^-$ & $(1,1\oplus3,\alpha+\frac{1}{2})^-$ & $(\bar{3},2,\alpha\!+\!\frac{1}{3})^-$ &  $\bar{Q}_L$ & $[X_3,X_4]|_{\alpha=-\frac{1}{2}}$\\
		A-2    & $S_{1}^{\dagger}$   &$d_R$ & $(1,1\oplus3,\alpha)^-$ & $(1,2,\alpha+\frac{1}{2})^-$ & $(\bar{3},1\oplus3,\alpha\!+\!\frac{1}{3})^-$ &  $\bar{Q}_L$ & $[X_3,X_4]|_{\alpha=0}$\\
		B-1     & $\widetilde{R}_2^{\dagger}$ &$\bar{Q}_L$   & $(1,2,\alpha)^-$ & $(1,1\oplus3,\alpha+\frac{1}{2})^-$ & $(3,1\oplus3,\alpha+\frac{1}{6})^-$ &   $d_R$ & $[X_3,X_4]|_{\alpha=-\frac{1}{2}}$\\
		B-2     & $\widetilde{R}_2^{\dagger}$ &$\bar{Q}_L$   & $(1,1\oplus3,\alpha)^-$ & $(1,2,\alpha+\frac{1}{2})^-$ & $(3,2,\alpha+\frac{1}{6})^-$ &   $d_R$ &   $[X_3,X_4]|_{\alpha=0}$\\
		C-1  & $U_{1}^{\dagger}$   &$\bar{u}_R$ & $(1,2,\alpha)^-$ & $(1,1\oplus3,\alpha+\frac{1}{2})^-$ & $(3,2,\alpha\!+\!\frac{2}{3})^-$ &  $Q_L$ & $[X_3,X_4]|_{\alpha=-\frac{1}{2}}$\\
		C-2    & $U_{1}^{\dagger}$   &$\bar{u}_R$ & $(1,1\oplus3,\alpha)^-$ & $(1,2,\alpha+\frac{1}{2})^-$ & $(3,1\oplus3,\alpha\!+\!\frac{2}{3})^-$ &  $Q_L$ &  $[X_3,X_4]|_{\alpha=0}$\\	
		D-1    & $\widetilde{V}_2^{\dagger}$ & $Q_L$   & $(1,2,\alpha)^-$ & $(1,1\oplus3,\alpha+\frac{1}{2})^-$ & $(\bar{3},1\oplus3,\alpha\!-\!\frac{1}{6})^-$ & $\bar{u}_R$ &$[X_3,X_4]|_{\alpha=-\frac{1}{2}}$\\
		D-2  & $\widetilde{V}_2^{\dagger}$   &$Q_L$& $(1,1\oplus3,\alpha)^-$ & $(1,2,\alpha+\frac{1}{2})^-$ & $(\bar{3},2,\alpha\!-\!\frac{1}{6})^-$ &  $\bar{u}_R$   & $[X_3,X_4]|_{\alpha=0}$\\	
		\hline\hline		
	\end{tabular}
	\caption{Same as in Table~\ref{tab:T1-i-a} but for the diagram T3-x depicted in Fig.~\ref{fig:4d2loop} and with the combination (a)-(A) shown in Fig.~\ref{fig:Z2}.}
	\label{tab:T3-x-a}   
\end{table*}

\bibliographystyle{apsrev4-1}
\bibliography{reference}

\end{document}